\documentclass[10pt,twocolumn,aps,pre,amsmath,amssymb,superscriptaddress]{revtex4-1}
\usepackage{times}
\usepackage{color}
\usepackage{graphicx}
\usepackage{bm}
\usepackage{amsmath}
\usepackage{amsfonts}
\usepackage{amsthm}
\usepackage{amssymb}
\usepackage{mathrsfs}
\usepackage{hyperref}
\begin{document}
\title{ Thermal transport  in out-of-equilibrium quantum harmonic chains}
\author{F. Nicacio}
\email{fernando.nicacio@ufabc.edu.br } 
\affiliation{ Centro de Ci\^encias Naturais e Humanas, 
              Universidade Federal do ABC, 09210-170, 
              Santo Andr\'e, S\~ao Paulo, Brazil }              
\author{A. Ferraro}
\affiliation{Centre for Theoretical Atomic, Molecular, and Optical Physics, 
             School of Mathematics and Physics, Queen's University, 
             Belfast BT7 1NN, United Kingdom}
\author{A. Imparato}
\affiliation{Dept. of Physics and Astronomy, University of Aarhus, Ny Munkegade, 
             Building 1520 - DK-8000 Aarhus C, Denmark}
\author{M. Paternostro}
\affiliation{Centre for Theoretical Atomic, Molecular, and Optical Physics, 
             School of Mathematics and Physics, 
             Queen's University, Belfast BT7 1NN, United Kingdom}
\author{F. L. Semi\~ao}
\affiliation{Centro de Ci\^encias Naturais e Humanas, 
             Universidade Federal do ABC, 
             Santo Andr\'e, 09210-170 S\~ao Paulo, Brazil}
\date{\today}
\begin{abstract}
We address the problem of heat transport in a chain of coupled quantum harmonic 
oscillators, exposed to the influences of local environments of various nature, stressing 
the effects that the specific nature of the environment has on the phenomenology 
of the transport process. 
We study in detail the behavior of thermodynamically relevant quantities such as 
heat currents and mean energies of the oscillators, establishing rigorous analytical 
conditions for the existence of a steady state, whose features we analyze carefully. 
In particular, we assess the conditions that should be faced to recover trends 
reminiscent of the classical Fourier law of heat conduction and highlight how such 
a possibility depends on the environment linked to our system. 
\end{abstract}
\maketitle
%
Understanding the transport properties in open systems in contact with several energy 
or particle baths represents a challenge for nonequilibrium physics.
Ideally, one would like to characterize and even calculate explicitly the statistics of 
the energy and particle currents, similarly to what can be done with observables in 
ensembles at equilibrium.
However the properties of the currents 
in out-of-equilibrium systems depends strongly on the bath properties, 
and on the characteristics of the system-bath coupling.
In this context, chains of oscillators have been extensively 
used as microscopic models for heat conduction and, in general, for
out-of-equilibrium systems~\cite{dhar,livi,hans12,hans14} 
to investigate the behavior of the thermal conductivity for different 
interaction potentials between the oscillators, different bath properties, 
or different system-bath couplings~\cite{livi,dhar}. 

The Fourier law of heat conduction implies that the heat current $J$
flowing throughout a system under a temperature gradient scales as
the inverse of the system size $L$, {\it i.e.}, $J\sim1/L$.
In the classical case, it is known that this law is violated in
1D homogeneous harmonic systems~\cite{rieder,hans12} where heat is carried by 
freely propagating elastic waves, while the current scales with the system size 
in presence of anharmonicity or disorder either in mass or in the coupling constant. 
Although the transport is anomalous 
($J\sim1/L^\alpha$, with $\alpha\ne1$) 
in these cases, the Fourier law is finally restored only 
in presence of a external substrate potential~\cite{dhar} or 
in the presence of a locally attached energy-conserving 
reservoirs for each oscillator~\cite{landi}. 

Quantum mechanically, a realistic description of a quantum medium for the transport 
of heat would imply the use of an explicitly open-system formalism and the introduction 
of system-environment interactions. In this context, it is interesting to identify 
the conditions, if any, under which heat transport across a given quantum system 
can be framed into the paradigm of Fourier law. Finding a satisfactory answer to 
this question is certainly not trivial, in particular in light of the ambiguities 
that the validity of Fourier law has encountered even in the 
classical scenario~\cite{rieder,hans12}. 
%

In the quantum scenario, significant studies are embodied in the work by 
Martinez and Paz~\cite{martinez}, who show the emergence of the three laws of thermodynamics in 
arbitrary networks evolving under a quantum Brownian master equation. 
This has been applied in Ref.~\cite{freitas} to show that the heat transport in this system is anomalous. 
Assadian {\it et al.}~\cite{assadian}, on the other hand, have addressed a chain of oscillators 
described by a Lindblad master equation, finding that a Fourier-like dependence on the 
system size can be observed for very long harmonic chains in the presence of dephasing.
Our work is also concerned with heat transfer in harmonic chains but the physical 
environment surrounding the chain is what differentiates our work from the ones previously cited. 
Basically, we include the possibility of establishing a temperature gradient 
using purely diffusive reservoirs, something not yet considered in the literature. 
Additionally, we further explore the effect of having the chain members 
locally attached to regular thermal baths at different temperatures. 

In this paper, we contribute to such research efforts by studying a general quadratic model for 
the dynamics of the system that, in turn, are affected by individual 
thermal reservoirs and exposed to the temperature gradient generated by all-diffusive 
environments. Our approach is able to pinpoint the origins of the specific 
energy distributions observed by varying the operating conditions of the system and 
thus identify the role played, respectively, by the diffusive and thermal reservoirs 
in the process of heat transport. We find working configurations that deviate 
substantially from the expectations arising from Fourier law and single out scenarios 
that are strictly adherent to such a paradigm, thus remarking the critical role played 
by the nature of the environment affecting the medium in the establishment of the actual 
mechanism for heat transport. 

The remainder of this paper is organized as follows. 
In Sec.~\ref{TAN} we introduce the formalism used to address the dynamics of the system.  
The general scenario addressed in our investigation is described in Sec.~\ref{TSAID}, 
while the thermodynamic properties and phenomenology of heat currents 
are analyzed in Sec.~\ref{AOHCATC}. Section~\ref{ATPC} is devoted to 
the analysis of a few significant cases that help us addressing the deviations from
(and adherences to) Fourier law.  
Finally, Sec.~\ref{conc} is devoted to the conclusions.

\section{Tools and notation}   \label{TAN}  
In this section we will consider a large class of systems with a generic number 
of degrees of freedom $n$. Let us define the operator
\begin{equation}
\hat x = (\hat q_1,...,\hat q_n, \hat p_1,...\hat p_n)^\dag , 
\end{equation}
which is the column vector composed by $n$ generalized coordinates together
with $n$ canonical conjugate momenta. 
%
It is possible to express the canonical 
commutation relations involving coordinates and momenta compactly as 
$[\hat x_j , \hat x_k ] = i \hbar \, \mathsf J_{jk}$ with ${\sf J}_{ij}$ 
the elements of the symplectic matrix
\begin{equation}                                                                         \label{comm}
 \mathsf J = 
\left( \!\! \begin{array}{rc} 
       {\bf 0}_n   & \mathsf I_n  \\
      -\mathsf I_n & {\bf 0}_n 
       \end{array}
\!\! \right).
\end{equation}
Here $\mathsf I_n$ and ${\bf 0}_n$ are the $n$ dimensional identity and zero matrix, 
respectively.
In the remainder of these notes, we will be dealing with quadratically 
coupled harmonic oscillators. In this scenario, the use of first 
and second moments of $\hat x$ provides a powerful tool for the description of 
the physically relevant quantities involved in the evolution of the system. 
We thus introduce the mean value (MV) vector 
$\langle \hat x \rangle_t  =  {\rm Tr}\left[ \hat x \hat\rho (t) \right]$ 
and the covariance matrix (CM) ${\bf V}$ of elements
\begin{equation}                                                                         \label{cmdef}
\mathbf V_{\! jk} (t)  =  
\tfrac{1}{2} {\rm Tr}\left[ 
                           \left\{ \hat x_j - \langle \hat x_j \rangle_t , 
                           \hat x_k - \langle \hat x_k \rangle_t \right\}
                           \hat\rho(t)
                     \right].  
\end{equation}

The focus of our work will be the study of a nearest neighbor-coupled 
harmonic chain whose 
$n$ elements are in contact with (individual) local reservoirs at finite 
temperature. 
The evolution of the chain can thus be described, in general, through the 
Lindblad master equation  
\begin{equation}                                                                         \label{lindblad}
\frac{d \hat\rho }{d t}  = 
         -\frac{i}{\hbar} [\hat H,  \hat\rho ] 
       - \frac{1}{2\hbar} \! \sum_{m}
             ( 
                    \{ \hat{L}_m^\dag \hat{L}_m , \hat\rho  \}
                    - 2 \hat{L}_m \hat\rho \hat{L}_m^\dag   )
\end{equation} 
with the following general form of a quadratic Hamiltonian and 
linear Lindblad operators
\begin{equation}                                                                         \label{ham-lind}
\hat H = \frac{1}{2} \hat x \cdot \mathbf H \hat x + 
\xi \cdot \mathsf J \hat x + H_0,~~
\hat L_m = \lambda_m  \cdot \mathsf J \hat x + \mu_m,
\end{equation}
where $\mathbf H$ is the adjacency matrix of the Hamiltonian, 
$\xi \in \mathbb R^{2n}$ 
is a column vector encompassing possible position and momentum 
displacements, 
$H_0 \in \mathbb R$ represents a possible energy offset, 
and $\lambda_m  \in \mathbb C^{2n}$ contains the coupling strengths between 
a given element of the chain and the respective reservoir. 
Finally, $\mu_m \in \mathbb C$  are constants \cite{endnote2}.
Using such a general description of the coherent and incoherent part 
of the evolution, 
we can straightforwardly work out the dynamical equations of motion for both 
$\langle\hat x\rangle_t$ and the CM 
{by calculating their time derivative and using 
the state evolution provided by Eq.~(\ref{lindblad})} as~\cite{nicacio}
\begin{equation}                                                                         \label{mvcm}
\frac{d\langle \hat x \rangle_t}{d t}  = 
\xi - \eta + { \bf \Gamma } \langle \hat x \rangle_t ,~~\frac{d \mathbf V}{d t}  = 
{ \bf \Gamma }\mathbf V + \mathbf V {\bf \Gamma}^\top + {\bf D} , 
\end{equation}
where we have introduced $\eta= \sum_{m } {\rm Im}(\mu_m^\ast \, \lambda_m)$ and
\begin{equation}                                                                         \label{dynmat}
{\bf \Gamma } = \mathsf J \mathbf H - {\rm Im} {\bf \Upsilon} \mathsf J,~~~~{\bf D} =  \hbar \, {\rm Re}{\bf \Upsilon},
\end{equation}
which are defined in terms of  the {\it decoherence} matrix 
${\bm \Upsilon}= \sum_{m} \lambda_m \lambda_m^\dagger$. 
By definition, we have 
${\rm Im} {\bf \Upsilon}^\top = - {\rm Im} {\bf \Upsilon}$ and 
${\bf D} = {\bf D}^\top \ge 0$. 

For time independent problems, Eqs.~(\ref{mvcm}) can be solved exactly as
\begin{equation}                                                                         \label{mvcmsol}
\begin{aligned}
\langle \hat x \rangle_t & = 
{\rm e}^{{\bf \Gamma} t }\langle \hat x \rangle_0 + 
{\bf \Gamma}^{-1} \left( {\rm e}^{{\bf \Gamma} t } - 
                         \mathsf I_{2n} \right)( \xi - \eta),  \\
{\bf V}(t) & = {\rm e}^{{\bf \Gamma} t} \, {\bf V}\!_0  \, 
            { \rm e}^{ {\bf \Gamma}^{\! \top} t } +
 \int_0^t \! dt^\prime \, 
               {\rm e}^{{\bf \Gamma} t^\prime} \, 
                  {\bf D}  \,
               {\rm e}^{{\bf \Gamma}^{\! \top} t^\prime}
\end{aligned}
\end{equation}
with $\langle \hat x \rangle_0$ and ${\bf V}\!_0$ 
the MV and CM of the initial state, respectively.  
%
%
%
The steady-state (or fixed-point) solutions of such equations can be found by imposing 
%
${d}\mathbf V/{d t} = {d}\langle \hat x \rangle_t/{d t}  = 0$, 
which are equivalent to the conditions 
(in what follows, the  subscript $\star$ will be used to indicate steady-state values)
\begin{equation}                                                                         \label{mvcmss}
\langle \hat x \rangle_{\star} = - { \bf \Gamma }^{-1}( \xi -\eta),~~{ \bf \Gamma }\mathbf V_{\!\star} + 
\mathbf V_{\!\star} {\bf \Gamma}^\top + {\bf D} = 0.  
\end{equation}
The equation satisfied by ${\bf V}_{\!\star}$ is of the 
stationary Lyapunov form~\cite{horn} that, under the conditions above, 
admits a unique positive-definite solution iff 
the eigenvalues of $\bf \Gamma$ have positive real parts. In this case, we find
\begin{equation}                                                                         \label{cmss}
{\bf V}_{\!\star} = \lim_{t \to \infty }{\bf V}(t)  =              
 \int_0^\infty \! \! dt \, 
               {\rm e}^{{\bf \Gamma} t} \, 
                  {\bf D}  \,
               {\rm e}^{{\bf \Gamma}^{\! \top} \! t}  \, .
\end{equation}
For-time dependent $\bf H$, $\xi$, $\eta$, and 
$\lambda_m$, the form of $ \langle \hat x \rangle_{\star} $ 
in Eq.~(\ref{mvcmss}) is no longer valid 
and the conditions over the Lyapunov equation for ${\bf V}_{\!\star}$ 
must hold at each instant of time. 

Note that, in order to deduce Eqs.~(\ref{mvcm}), (\ref{mvcmsol}), and (\ref{cmss}), 
we did not need to make any assumption on the initial state 
of the system but only use the quadratic and linear structure 
of Eq.~(\ref{ham-lind}) and (\ref{lindblad}), respectively. 
The rest of our analysis will focus on the dynamics of 
thermodynamically relevant quantities such as 
currents and energy. 

\section{The System and its dynamics} \label{TSAID} 
\begin{figure}[t!]                                                      
\includegraphics[width=8cm]{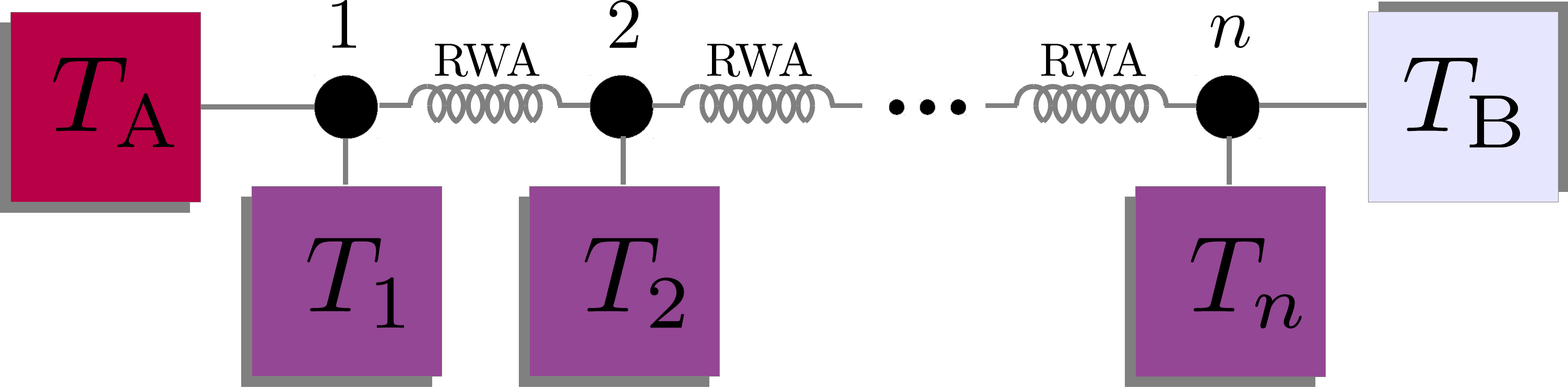} 
\caption{(Color online) Schematic representation of the system. 
A chain of coupled harmonic oscillators interacting according 
to an RWA model.}                                                                        \label{fig1chain}                            
\end{figure}

We can now start analyzing explicitly the system that we have in mind.  
We consider the chain of oscillators depicted in Fig.~\ref{fig1chain}, 
each interacting with its own thermal reservoir at temperature $T_k, \, k=1,...,n $. 
The Hamiltonian of the chain arises from the application of the rotating-wave 
approximation on a nearest-neighbour Hooke-like coupling model, which gives us
\begin{equation}                                                                         \label{hamsys1}
\hat H =  \hbar\omega \sum_{j = 1}^n \hat{a}_j^\dag \hat{a}_j + 
            2 \hbar \Omega \sum_{j = 1}^{n - 1} 
            (\hat{a}_j^\dag \hat{a}_{j+1} + \hat{a}_{j+1}^\dag \hat{a}_j), 
\end{equation}
where $\omega$ and $\Omega$ are the frequency of the oscillators and their mutual 
coupling rate respectively, and $\hat a_j=(\hat q_j + i \hat p_ j)/{ \sqrt{2\hbar} }$ 
is the creation operator of the $j^{\rm th}$ oscillator. 
Using the notation introduced before, $\hat H$ can be written as in 
Eq.~(\ref{ham-lind}) with $\xi = H_0 = 0 $ and the adjacency matrix 
$\mathbf H = {\pmb H} \oplus {\pmb H}$, where 
\begin{equation}                                                                         \label{hamsys2}
{\pmb H}_{\! j k} = \omega \, \delta_{j k} + 
                    \Omega \, (\delta_{j\, k+1} + \delta_{j \, k-1} )
\end{equation}
and $\delta_{j k}$ is the Kronecker symbol. 

The coupling between a given oscillator and the respective thermal 
reservoir is described by the Lindblad operators \cite{endnote3}
\begin{equation}                                                                         \label{lind22}
\hat L_k  = \sqrt{\hbar\zeta_k (\bar N_k + 1)} \, 
\hat a_k ,~~\hat L_k' = \sqrt{\hbar\zeta_k \bar N_k} \, \hat a_k^\dagger,
\end{equation}
where $\zeta_k \ge 0$ is the bath-oscillator coupling and 
$\bar N_k$ is the mean occupation number of the reservoirs at 
temperature $T_k$. 
This choice allows us to make the identifications 
\begin{eqnarray}                                                                         \label{lindbath}
{\lambda}_k &=&  \sqrt{\frac{\zeta_k}{2} (\bar N_k + 1)} \, 
( \underbrace{ 0 ,..., 0 }_{ k-1 } , 
  i, 
  \underbrace{ 0 ,..., 0 }_{ n-1 } , 
  -1,
  \underbrace{ 0 ,..., 0 }_{ n-k } )^\top, \nonumber \\
\lambda'_{k} &=& \sqrt{\frac{\zeta_k}{2} \bar N_k} \,
( \underbrace{ 0 ,..., 0 }_{ k-1 } , 
  -i, 
  \underbrace{ 0 ,..., 0 }_{ n-1 } , 
  -1,
  \underbrace{ 0 ,..., 0 }_{ n-k } )^\top. 
\end{eqnarray}

We now make the explicit assumption that the first and last oscillator 
in the chain are also affected by two additional reservoirs, 
which we label ${\rm A}$ and ${\rm B}$, having temperatures 
$T_{\rm A} \ge T_{\rm B} \gg T_k$. 
This allows us to approximate $\bar N_l + 1 \approx \bar N_l$ 
for $ l = {\rm A},{\rm B} $. 
Therefore, such baths contribute with 
\begin{equation}                                                                         \label{lindexbath}
\begin{aligned}
\lambda'_{\rm A}  &= \lambda^\ast_{\rm A} = 
\sqrt{\frac{\zeta_{\rm A}}{2} \bar N_{\rm A} } \, 
(  i, 
  \underbrace{ 0 ,..., 0 }_{ n-1 } , 
  -1, 
  \underbrace{ 0 ,..., 0 }_{ n-1 })^\dagger , \\
\lambda'_{\rm B}  &= \lambda^\ast_{\rm B} = 
\sqrt{\frac{\zeta_{\rm B}}{2} \bar N_{\rm B} } \, 
( \underbrace{ 0 ,..., 0 }_{ n-1 } , 
  i, 
  \underbrace{ 0 ,..., 0 }_{ n-1 }, 
  -1 )^\dagger.
  \end{aligned}
\end{equation}

We are now in a position to give some motivations for the specific 
choice of the system to study. 
In a realistic scenario, the impossibility to achieve full isolation 
leads one to take into account the external and uncontrollable influences 
from the environment over the evolution of a system. 
In our case such disturbances are represented by the 
$n$ thermal reservoirs attached to each oscillator of the chain. 
On the other hand, as we aim at studying heat transport across the system, 
we need to set a temperature gradient, which is 
imposed, in our setting, by the external end-chain baths. As such gradient is 
supposed to be the leading mechanism for the transport process, it  
is reasonable to assume that $T_{A,B}$ are the largest 
temperatures across the system.    

%
%

We can now go back to the formal description of the system and write 
the decoherence matrix as ${\bf \Upsilon} = {\bf \Upsilon}^{({\rm A})} + {\bf \Upsilon}^{({\rm B})} 
                + \sum_{k = 1}^{n} {\bf \Upsilon}^{(k)}$
with 
\begin{equation}                                                                         \label{decsys2} 
\begin{aligned}
&{\bf \Upsilon}^{(k)} = 
                      \lambda_k \lambda_k^\dagger  + 
                      \lambda_k' \lambda_k'^\dagger~~~(k=1,\dots,n),  \\
&{\bf \Upsilon}^{( l )} = 
2 \, {\rm Re}  (\lambda_{l } \lambda_{ l}^\dagger )~~~~~~~~~(l={\rm A},{\rm B}).
\end{aligned}
\end{equation}
As the contribution given by the reservoirs ${\rm A}$ and ${\rm B}$ to the dynamics 
is all in the matrix $\mathbf D$ of Eq.~(\ref{dynmat}), 
we refer to them as {\it all-diffusive}.  

In order to simplify our analysis without affecting its generality, 
we now take $\zeta_k = \zeta$. 
From Eq.~(\ref{hamsys2}) and the expression found for ${\bm\Upsilon}$, 
we rewrite (\ref{dynmat}) as
\begin{equation}                                                                         \label{dynsys}
{ \bf \Gamma }  = -
\begin{pmatrix}
 \frac{\zeta}{2} \, \mathsf I_{n} & - {\pmb H}\\
  {\pmb H} &  \frac{\zeta}{2} \, \mathsf I_{n} 
\end{pmatrix},
%
\,\,\, 
{\bf D} =  \frac{\hbar\zeta}{2} \, \mathsf I_{2n} +  
            {\pmb D} \! \oplus \! {\pmb D}
\end{equation}
with ${\pmb D} = \hbar\zeta \, 
{\rm Diag}( \frac{\zeta_{\rm A}}{\zeta} \bar{N}_{\rm A} + \bar{N}_{1} , 
            \bar{N}_{2},..., \bar{N}_{n-1},
             \frac{\zeta_{\rm B}}{\zeta}\bar{N}_{\rm B} + \bar{N}_{n}) $.
This allows us to achieve the following expression for the CM using (\ref{mvcmsol})
\begin{equation}                                                                         \label{cmsolt}
\!{\bf V}(t)  =  {\rm e}^{ {\bf \Gamma} t} \, {\bf V}\!_0  \, 
              { \rm e}^{ {\bf \Gamma}^{\! \top} t } +
              \tfrac{\hbar}{2} \left( 1 - {\rm e}^{-\zeta t }\right)\mathsf I_{2n} +
              {\bf U}^\dagger \,( {\bf I} \oplus {\bf I}^\ast ) \,  {\bf U} 
\end{equation}
with
${\bf I} =  \int^t_0 \! \! dt'  \, \, 
{\rm e}^{-\zeta t'} {\rm e}^{-i {\pmb H} t'} {\pmb D }\, {\rm e}^{i {\pmb H} t'}$ 
and $\bf U$ defined in (\ref{matrixU}). 
%
Following the lines given in the Appendix and integrating ${\bf I}$ by parts, 
it is then possible to show that 
\begin{equation}                                                                         \label{cmsyst}
\begin{aligned}
\!\!\!\!\!\! {\bf V}(t)  & =   {\rm e}^{ {\bf \Gamma} t} \, {\bf V}\!_0  \, 
                          { \rm e}^{ {\bf \Gamma}^{\! \top} t } 
+\tfrac{\hbar}{2} \left( 1 - {\rm e}^{-\zeta t }\right)\mathsf I_{2n}                  \\ 
&+ {\bf O}  \! \oplus  \!  {\bf O} 
          \left( \!\! \begin{array}{rr}
                      {\bf O} {\pmb D} {\bf O} \circ {{\rm Re} \bf L} & 
                    - {\bf O} {\pmb D} {\bf O} \circ {{\rm Im} \bf L}   \\                         
                      {\bf O} {\pmb D} {\bf O} \circ {{\rm Im} \bf L} & 
                      {\bf O} {\pmb D} {\bf O} \circ {{\rm Re} \bf L}    
                 \end{array}   \!\!                                        \right) 
{\bf O} \! \oplus \! {\bf O}, 
\end{aligned}
\end{equation}
with $\circ$ the symbol for a Hadamard matrix product~\cite{horn} 
and the matrices ${\bf O}$ and ${\bf L}$ having the elements   
\begin{equation}                                                                        \label{matrixOL}
\begin{aligned}
&{\bf O}_{kl}=\sqrt{\frac{2}{n+1}}\sin\left(\frac{kl\pi}{n+1}\right), \\
&{\bf L}_{jk}= \frac{1 - {\rm e}^{-[\zeta + i (\nu_j -\nu_k)] t } }
{ \zeta + i ( {\nu_j -\nu_k  }) } 
\end{aligned}
\end{equation}
with $\nu_m = \omega + 2 \Omega \cos(\tfrac{m \,\pi}{ n + 1})$ ({\it cf.} the Appendix). 
The steady-state form of such solution can be found as illustrated 
in the previous section, which yields 
\begin{equation}                                                                         \label{cmsssys}
\begin{aligned}
&{\bf V}_{\!\star}  = 
\frac{\hbar}{2}\mathsf I_{2n} \,\, + \\
&{\bf O}  \! \oplus  \!  {\bf O} 
\begin{pmatrix}
                   {\bf O} {\pmb D} {\bf O} \circ {{\rm Re} \bf L_\star} & 
                 - {\bf O} {\pmb D} {\bf O} \circ {{\rm Im} \bf L_\star}   \\                         
                   {\bf O} {\pmb D} {\bf O} \circ {{\rm Im} \bf L_\star} & 
                 {~~}{\bf O} {\pmb D} {\bf O} \circ {{\rm Re} \bf L_\star}     
\end{pmatrix}
{\bf O} \! \oplus \! {\bf O}
\end{aligned}
\end{equation}
with 
${{\bf L}_\star}_{jk} := \lim_{t \to \infty} {\bf L}_{jk} 
=1/ [\zeta + i ( {\nu_j -\nu_k  })]$.
Remarkably, Eq.~\eqref{cmsssys} is the CM of a vacuum state corrected 
by terms whose origin is entirely ascribed to the presence of the reservoirs. 
Furthermore, the nullity of the diagonal elements of ${{\rm Im} \bf L}_\star$ 
guarantees that, in the long-time limit, each oscillator is in a thermal state. 
If the reservoirs connected to the elements of the chain have all the same 
temperature 
(so that $\bar{N}_k=\bar{N},~\forall k=1,\dots,n$),
Eq.~\eqref{cmsssys} can be cast into the form
\begin{equation}                                                                         \label{cmsssys4}
\begin{aligned}
&{\bf V}_{\!\star}  =   
\hbar(\bar N + \tfrac{1}{2}) \mathsf I_{2n} \,\, + \\ 
& {\bf O}  \! \oplus  \!  {\bf O} 
          \left( \!\! \begin{array}{rr}
                   {\bf O} {\pmb D} {\bf O} \circ {{\rm Re} \bf L_\star} & 
                 - {\bf O} {\pmb D} {\bf O} \circ {{\rm Im} \bf L_\star}   \\                         
                   {\bf O} {\pmb D} {\bf O} \circ {{\rm Im} \bf L_\star} & 
                   {\bf O} {\pmb D} {\bf O} \circ {{\rm Re} \bf L_\star}     
                      \end{array}   \!\!                           \right) 
{\bf O} \! \oplus \! {\bf O},
\end{aligned} 
\end{equation}
with 
${\pmb D} := \hbar \, {\rm Diag}
( \zeta_{\rm A} \bar{N}_{\rm A},0,...,0,
  \zeta_{\rm B} \bar{N}_{\rm B} )$. This is the CM of a thermal equilibrium state at 
temperature $T$ for the $n$ oscillators plus corrections due to the 
all-diffusive reservoirs. 
In the Appendix, we analyze in details some aspects of the structure of 
the CM in (\ref{cmsssys}) and (\ref{cmsssys4}).  

\section{Analysis of heat current across the chain} \label{AOHCATC}     
In a thermodynamical system ruled by Hamiltonian $\hat H$ and described by the density 
matrix $\hat \rho$, the variation of the internal energy is associated with 
work and heat currents. In fact
\begin{equation}                                                                         \label{dHdt}
\frac{ d }{d t} \langle \hat H \rangle = 
{\rm Tr}\left(\hat\rho \frac{\partial \hat H }{\partial t}\right) + 
{\rm Tr} \left( \frac{d \hat\rho }{d t} \hat H \right).
\end{equation}
While the first term in the right-hand side is associated with the work performed on/by 
the system in light of the time-dependence of its Hamiltonian, the second term accounts 
for heat flowing into/out of the system itself. 
As the Hamiltonian of our problem is time-independent, any change in the mean energy of 
the chain should be ascribed to the in-flow/out-flow of heat. 
By inserting the right-hand side of Eq.~\eqref{lindblad} in place 
of $d\hat\rho/dt$ above, we find 
\begin{equation}                                                                         \label{totcurr1}
\mathcal J = {\rm Tr} \left( \frac{d \hat\rho }{d t} \hat H \right) = 
                \sum_{k }  \mathcal J_k  
\end{equation}
with $\mathcal J_k  =  \frac{1}{2\hbar}   
                    \langle  
                    2 \hat{L}_k \hat H \hat{L}_k^\dag  - \{\hat H, \hat{L}_k^\dag \hat{L}_k \} \rangle$
the heat current induced by the $k^{\rm th}$ Lindblad operator. 
Physically speaking, Eq.~(\ref{totcurr1}) shows that the total 
heat current in the system is formed by the net result of currents due to each reservoir. 
This largely enriches the phenomenology of heat propagation and thermalization in our system, 
especially compared to usual previous settings~\cite{assadian} .

This expression can be specialized to the case of 
the system addressed in Sec.~\ref{TSAID} to give ({\it cf.} the Appendix) 
\begin{eqnarray}                                                                         \label{partialcurr}
&&\mathcal J_k   =   \\
&& {\rm Tr} \left[\frac{\hbar}{2}  
                {\bf H } \,   {\rm Re} ( \lambda_k \lambda_k^\dag  ) 
  {- }{\bf H} 
           \left( {\bf V} + 
           \langle \hat x \rangle_{t} \langle \hat x \rangle_{t}^\top 
           \right) \mathsf J \, 
           {\rm Im} ( \lambda_k \lambda_k^\dag) 
           \right]. \nonumber 
\end{eqnarray}
Summing over all Lindblad operators,
the total current reads
\begin{equation}                                                                         \label{totcurr2}
\mathcal J  =  \frac{1}{2}  
{\rm Tr} \left[ {\bf H } \, {\bf D } \right] 
- {\rm Tr} \left[ {\bf H} 
                   \left( {\bf V} + 
                          \langle \hat x \rangle_{t} \langle \hat x \rangle_{t}^\top 
                   \right) \mathsf J \, 
                   {\rm Im}{\bf \Upsilon} 
           \right].
\end{equation}
The first term is the diffusive part of the current and is constant in time 
if the set of $\lambda_k$'s does not depend on time explicitly. 

\begin{figure}[t!]                                                      
\includegraphics[width=8cm]{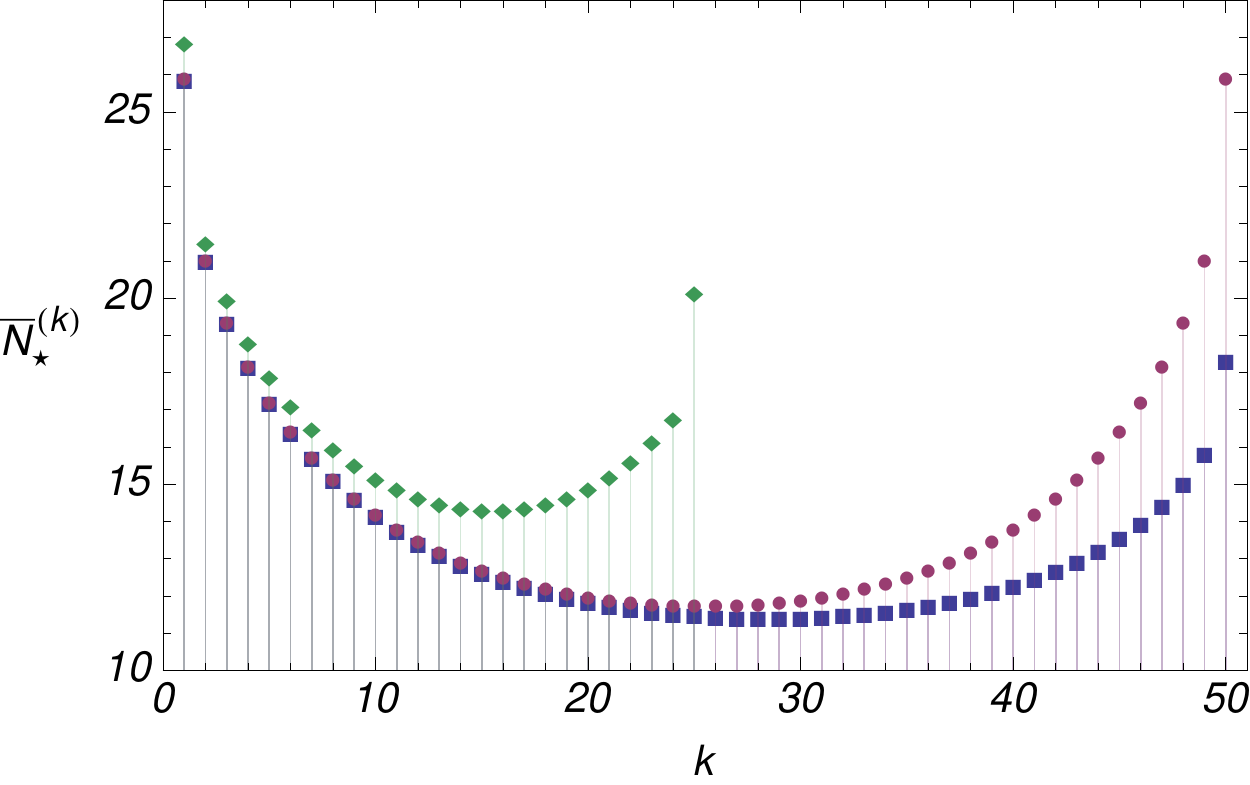} 
\caption{ 
(Color online) Distribution of mean occupation numbers for the 
elements of a chain of two lengths. 
The (green) diamond-shaped points correspond to a chain of $ n =  25 $ with 
$\bar N_{\rm A}  =  2 \bar N_{\rm B} = 10 \bar N_k = 100$. The (blue) square points
are for $ n =  50 $ with  $\bar N_{\rm A}  =  2 \bar N_{\rm B} = 10 \bar N_k = 100$. Finally, 
the (violet) dots are for $ n =  50 $ with $\bar N_{\rm A} =   \bar N_{\rm B} = 10 \bar N_k = 100$.  
The remaining parameters are $\Omega/\omega = 1/2$, 
$\zeta/\omega = \zeta_{\rm A}/\omega = \zeta_{\rm B}/\omega = 1/10$, $\hbar = 1$. 
For all the three plots k =1,..,n.}                                                                     \label{fig2therm} 
\end{figure}

The system's internal energy can be easily worked out to take the general form
\begin{equation}                                                                         \label{meansHtsys}
\begin{aligned}
\langle \hat H \rangle_t &= 
\frac{1}{2}{\rm Tr}\left[ {\bf H}\,{\bf  V}(t)  + 
{\bf H}\,\langle \hat x \rangle_t \langle \hat x \rangle_t^\top \right]\\ 
%
& =  \frac{1}{2}{\rm e}^{-\zeta t } \, {\rm Tr} (\mathbf H {\bf V}_{\!0}) +    
\frac{1}{2}{\rm e}^{-\zeta t}  
\langle \hat x \rangle_0 \cdot {\bf H} \langle \hat x \rangle_0   \\
& +  \hbar\omega \left( \frac{n}{2} + \!\sum_{k=1}^n \bar N_k \! + \!  
\frac{\zeta_{\rm A}}{\zeta} \bar N_{\rm A}  +  
\frac{\zeta_{\rm B}}{\zeta} \bar N_{\rm B} \right)
\left( 1 - {\rm e}^{-\zeta t} \right). 
\end{aligned}
\end{equation}
At the steady state, we can write 
\begin{equation}                                                                         \label{sshsys1}
\langle \hat H \rangle_\star  = 
\hbar\omega\left(\frac{\zeta_{\rm A}}{\zeta} \bar N_{\rm A} + 
\frac{\zeta_{\rm B}}{\zeta} \bar N_{\rm B}\right) 
+  \hbar\omega \sum_{k=1}^n \bar N_k   
+ \frac{1}{2}\hbar \omega n, 
\end{equation}
showing that the mean energy of the system does not depend on the coupling strength 
between the oscillators and is fully determined by the 
the interactions with the reservoirs. 
The contribution that each oscillator gives to the equilibrium energy 
in Eq.~(\ref{sshsys1}) is not uniform across the chain, as can be seen from Fig.~\ref{fig2therm} 
where we plot the mean occupation number of each oscillator
\begin{equation}                                                                         \label{meansssys}
\bar N^{(k)}_\star = {\rm Tr}(\hat a^\dag_k \hat a_k \hat\rho) 
                   = \frac{1}{\hbar}[{{\bf V}_{\! \star}}]_{kk} -1/2.   
\end{equation}
{Despite the individual contribution of each bath for the mean 
energy in (\ref{sshsys1}), the state of the chain is described by 
the CM (\ref{cmsssys4}), which encompasses the collective effects of all the reservoirs resulting from the mixing process 
effectively implemented by the inter	oscillator coupling}.
As for the current, one finds
\begin{equation}                                                                         \label{currsys}
\begin{aligned}
\mathcal J & = -\frac{\zeta}{2}{\rm e}^{-\zeta t } 
\, {\rm Tr} (\mathbf H {\bf V}_{\!0})    
-\frac{\zeta}{2}{\rm e}^{-\zeta t}  
\langle \hat x \rangle_0 \cdot {\bf H} \langle \hat x \rangle_0 \\
& + \hbar \omega \left(
\frac{\zeta n}{2} + \zeta \sum_{k=1}^n \bar N_k \! + \!  
\zeta_{\rm A} \bar N_{\rm A}  +  \zeta_{\rm B} \bar N_{\rm B} \right)
{\rm e}^{-\zeta t}
\end{aligned}
\end{equation}
with $\mathcal J_\star = 0$. 
As the current is a linear function of the matrix $\lambda_m \lambda_m^\dag $ 
[{\it cf.} Eq.~(\ref{partialcurr})], in order to interpret each term of the above 
equation and their contribution to the total current at the steady state, 
we break the total current into the three parts. The first two are time independent and read
\begin{equation}                                                                         \label{currdifsys}
\mathcal J^{(l)} = 
\frac{\hbar}{2} {\rm Tr} \left[ {\bf H} \, 
                                {\bf \Upsilon}^{(l)} 
                         \right] 
                                 =  \hbar \omega \zeta_{l} \bar{ N }_{l} 
~~~~~~~(l = {\rm A}, {\rm B}).
\end{equation}
The third one is
\begin{eqnarray}                                                                         \label{currdissys}
\mathcal J^{(k)} & = &
\frac{\hbar}{2}  
{\rm Tr} \left[ {\bf H }\, {\rm Re} {\bf \Upsilon}^{(k)} \right] 
- {\rm Tr} \left[ {\bf H} 
                   {\bf V}  \mathsf J \, 
                   {\rm Im} {\bf \Upsilon}^{(k)}  \right]  \\ 
& = & \hbar \omega \zeta (\bar N_k +1/2) \! 
- \!\zeta( \omega \, {\bf V}_{\!kk} + 
\Omega \,{\bf V}_{\! k-1 k}  + \Omega \,{\bf V}_{\! k k+1}   )\, . \nonumber 
\end{eqnarray}
For simplicity, we have omitted the explicit dependence on the initial conditions. 
The simple form attained in Eq.~(\ref{currdifsys}) is a consequence of Eq.~(\ref{decsys2}),
where the matrices 
${\bf \Upsilon}^{({\rm A})}$ and ${\bf \Upsilon}^{({\rm B})}$ 
are purely real.  
%
%
%
At the steady state, using Eqs.~(\ref{meansssys}) and (\ref{currdissys}), one finds 
\begin{equation}                                                                         \label{currssintsys}
\begin{aligned}
\mathcal J^{(1)}_\star &=
- \hbar\omega \zeta \left[\bar N^{(1)}_\star - \bar N_1\right]  
- \Omega \zeta \, {\bf V}_{\! \star 1 2 } ,                               \\ 
\mathcal J^{(k)}_\star &=
- \hbar\omega \zeta \left[\bar N^{(k)}_\star - \bar N_k\right]  
-  \Omega \zeta  ( {\bf V}_{\! \star k k+1} + {\bf V}_{\! \star k-1 k} ), \\ 
\mathcal J^{(n)}_\star &=
- \hbar\omega \zeta \left[\bar N^{(n)}_\star - \bar N_n\right]    
-  \Omega \zeta \,  {\bf V}_{\! \star n-1 n }.
\end{aligned}
\end{equation}
The above currents for each reservoir in the chain at the stationary state 
are plotted in Fig.~\ref{fig3curr}. 
It is possible to show, see the Appendix, that  
\begin{equation}                                                                         \label{zerocond}
{\bf V}_{\! \star j j+1} =  0,  \,\,\, (j = 1,..., n-1).
\end{equation}
Thus the currents in Eq.~(\ref{currssintsys}) are given exclusively by the 
difference between the mean occupation number of the reservoirs $\bar N_k$, 
and the mean thermal photon number of each oscillator $\bar N^{(k)}_\star$.  
That is, the heat currents within the system can be understood as the difference 
between the amounts of energy stored in a given reservoirs and that in 
the respective oscillator. 
The fact that $\mathcal J_\star = 0$ implies that all the internal currents ${\cal J}^{(k)}$ are 
constrained to sum up the (constant) value 
$- \left(\mathcal J^{({\rm A})} + \mathcal J^{({\rm B})}\right)$ independently 
on the length of the chain, 
which shows a clear violation of Fourier law of heat conduction. 
From Fig.~\ref{fig3curr} note that all the currents are negative showing also that 
the thermal energy stored in each oscillator, represented by $\bar N^{(k)}_\star$, 
is greater than the energy of its own reservoir, see Eq.(\ref{currssintsys}).  
\begin{figure}[htbp!]                                                      
\includegraphics[width=8cm]{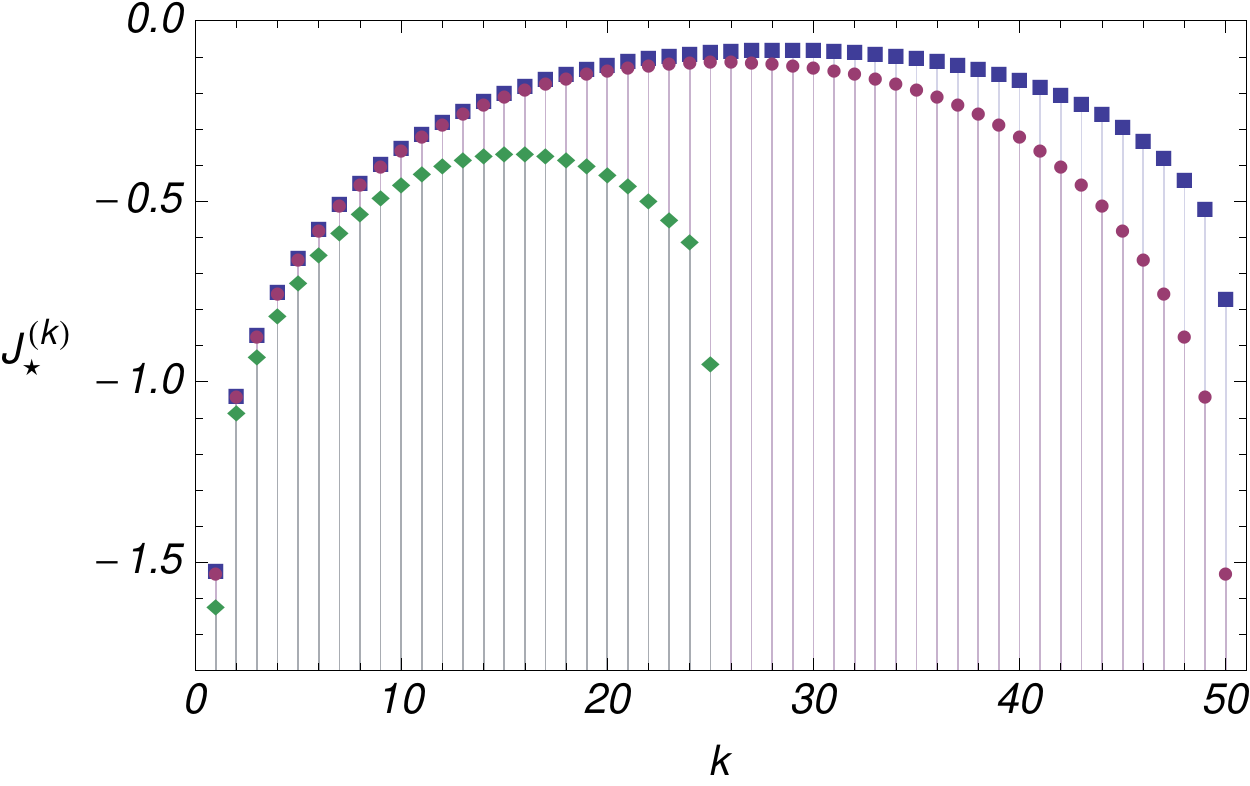} 
\caption{ 
(Color online) Currents across the chain. 
The (green) diamond-shaped points correspond to a chain of $ n =  25 $ with 
$\bar N_{\rm A}  =  2 \bar N_{\rm B} = 10 \bar N_k = 100$. The (blue) square points
are for $ n =  50 $ with  $\bar N_{\rm A}  =  2 \bar N_{\rm B} = 10 \bar N_k = 100$. Finally, 
the (violet) dots are for $ n =  50 $ with $\bar N_{\rm A} =   \bar N_{\rm B} = 10 \bar N_k = 100$.  
The remaining parameters are as in  Fig.~\ref{fig2therm}. 
}                                                                                        \label{fig3curr}                            
\end{figure}                                                          

\begin{figure}[t!]                                             
\includegraphics[width=8.5cm]{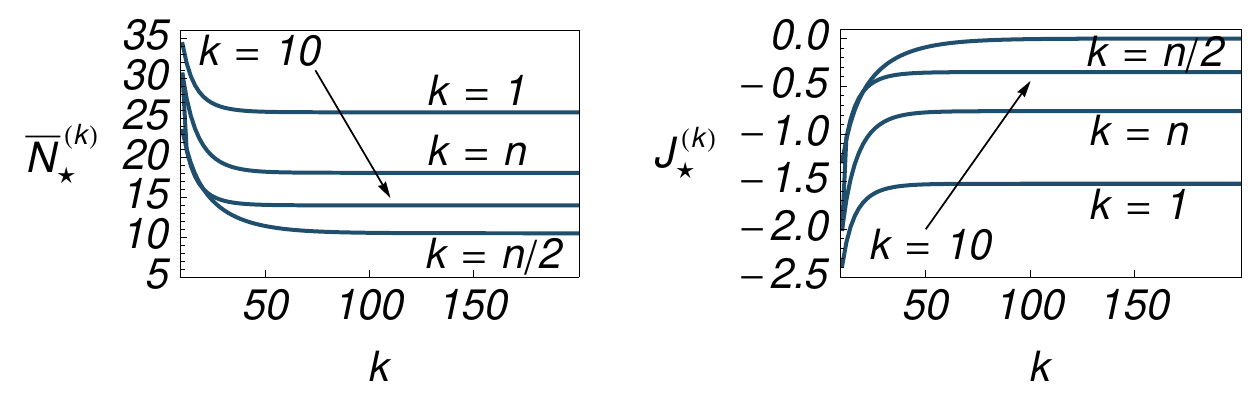} 
\caption{ 
(Color online) Mean value of energy (left) and currents (right) 
for the fixed-position oscillators [the first $( k = 1 )$ 
and the 10$^\text{th}$ ($ k = 10$ )], 
the midpoint (bulk) $k = n/2$,  and the last oscillator $k = n$. 
We consider $\bar N_{\rm A}  =  2 \bar N_{\rm B} = 10 \bar N_k = 100$; 
the remaining parameters are the same as in  Fig.~\ref{fig2therm}.}                       \label{fig4thercur1}                            
\end{figure}

The general behavior of both quantities against the length of the chain can be seen from
Fig.~\ref{fig4thercur1}. 
For this range of $n$ (chain lengths) the local occupation number decreases 
with $n$ while the current does the opposite. 
As far as we could check numerically, for $n \gg 1$, both quantities become independent of the length of the chain. 
However, due to the lack of an analytical proof we cannot assure that this is 
really the case in the thermodynamic limit. 
Obviously, the relation Eq.~(\ref{currssintsys}) among these quantities 
is valid for any chain length.   
Furthermore, 
any oscillator in the bulk of the chain, {\it i.e.}, 
any element identified by a label $k \sim n/2$, 
has the same occupation number of the reservoir attached to it. 
This is due to the fact that, by Eq.~(\ref{matrixOL}), 
${\bf O}_{k j} \approx \sqrt{2/n} \sin(j \pi/2)$  
and $\bar N^{(k)}_\star = \bar N_k$. 
Moreover, the individual currents, {\it i.e.}, 
the currents due to the local standard thermal baths, are null according to Eq.~(\ref{currssintsys}). 
This implies that in this limit the local baths play no major role in the thermalization of the bulk oscillators. 
In this respect, our results share some similarity with the classical approach 
of the problem, where a chain of harmonic oscillators is 
attached to just two thermal baths at its ends. 
In fact, in the work by Reider {\it et al.}~\cite{rieder}, 
the bulk oscillators attains a constant temperature as in our case 
when the limit $n\gg 1$ is considered.

%

\section{Application to paradigmatic cases} \label{ATPC}        
In this section we use the formalism and results illustrated so far to analyze the 
transport of heat in a few paradigmatic examples, all encompassed by the general 
treatment of the problem provided above. 
In particular, we want to study the effects caused by the presence of end-chain 
all-diffusive reservoirs, something not yet explored in this context before. 
To this end, we now consider different variations on the basic setup described on 
Fig.~{\ref{fig1chain}}. 
For example, we vary the distribution of reservoirs, 
the coupling mechanism among the oscillators and include dephasing. 

\subsection{Case I: Ordinary Baths } \label{caI}                                  
\begin{figure}[htbp!]                                                      
\includegraphics[width=8cm]{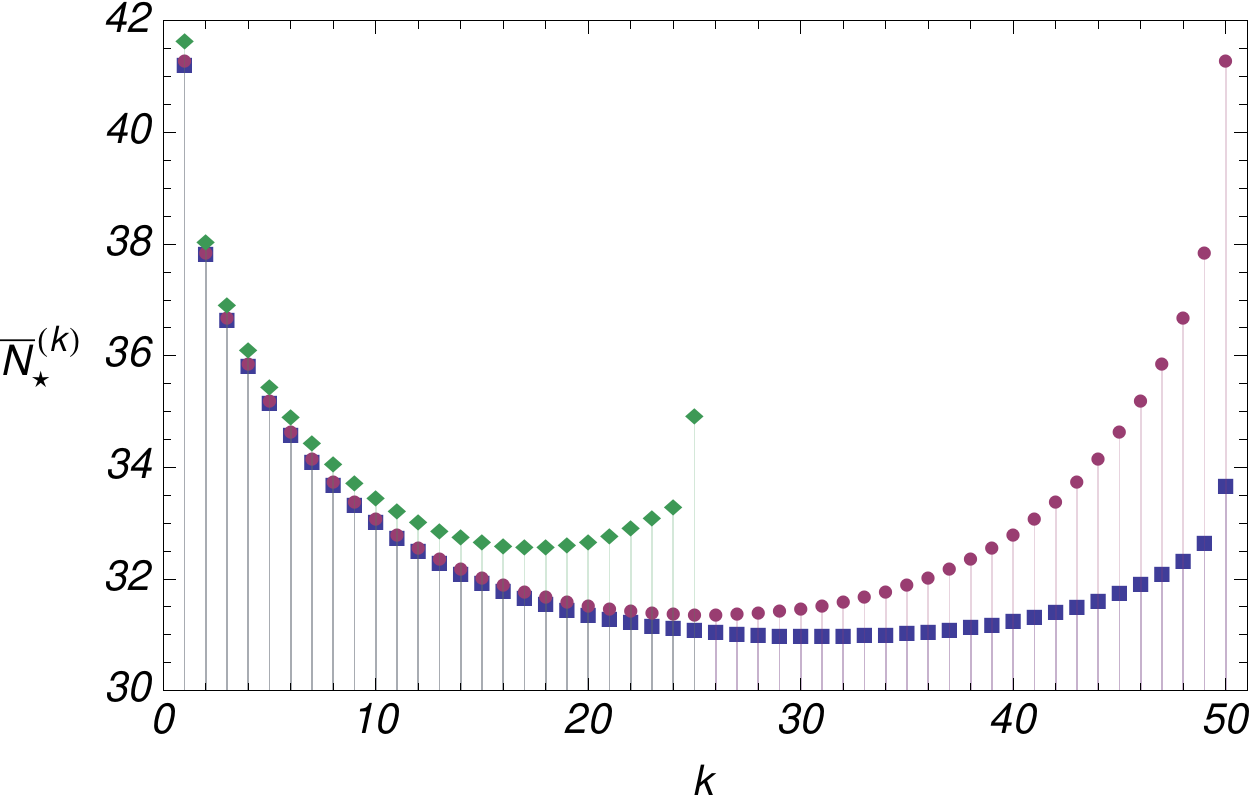} 
\caption{ 
(Color online) Distribution of mean occupation numbers for the 
elements of a chain without the diffusive reservoirs. 
Diamonds (green): chain with $ n =  25 $ oscillators and  
temperatures of the baths given in terms of 
$\bar N_{1}  =  2 \bar N_{n} = 100$ and $\bar N_{k} = 30$;  
Squares (blue): chain with $ n =  50 $  oscillators and  
$\bar N_{1}  =  2 \bar N_{n} = 100$ and $\bar N_{k} = 30$;  
Circles (violet): $ n =  50 $  and  
$\bar N_{1}  =  \bar N_{n} = 100$ and $\bar N_{k} = 30$.  
The remaining parameters are the same as in Fig.~\ref{fig2therm} 
except for $\zeta_{\rm A}$ and $\zeta_{\rm B}$, which taken to be zero and in 
all cases $  k = 2,...,n-1$.}                                                              \label{fig5therm}                            
\end{figure}

\begin{figure}[htbp!]                                                      
\includegraphics[width=8cm]{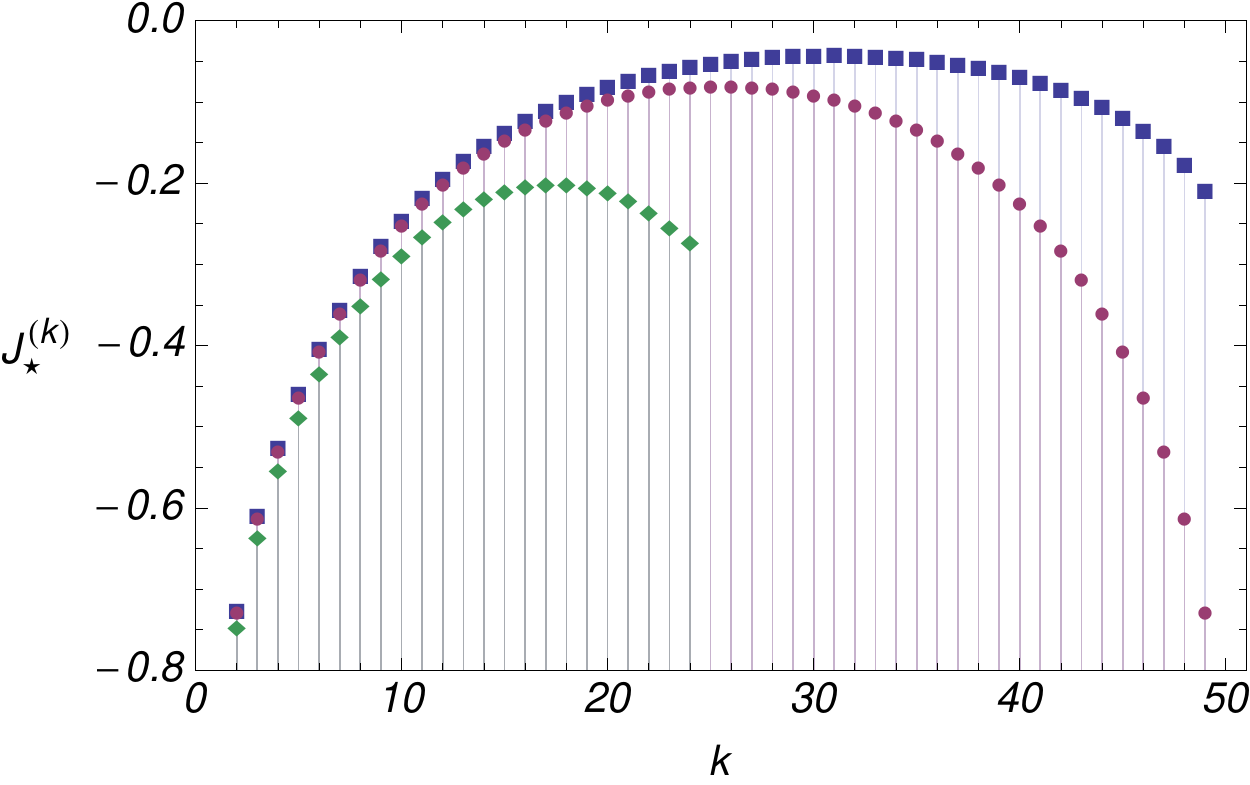} 
\caption{ 
(Color online) Currents across the chain without the diffusive reservoirs. 
The color-symbol code is the same as in Fig.~\ref{fig5therm}, 
although we have taken $k = 2,...,n-1$, {\it i.e.}, 
we have excluded the positive currents from the first and last reservoir.}                         \label{fig6curr}
\end{figure}

In order to establish a benchmark to evaluate the role played by the diffusive baths,
we start the analysis by taking $\zeta_{\rm A} = \zeta_{\rm B} = 0$ in (\ref{lindexbath}). 
%
%
%
The corresponding steady-state CM is as in Eq.~(\ref{cmsssys}) with 
\begin{equation}
{\pmb D} = \hbar \zeta \, {\rm Diag}({\bar N_1,...,\bar N_n}),
\end{equation}
while the steady-state energy and the current are given 
by Eq.~(\ref{sshsys1}) and Eq.~(\ref{currsys}), respectively. 
%
%
%
%
%
In order to remain as close as possible to the system discussed in the previous section,
we take $T_1 > T_k > T_n$ for $  1 < k <  n$. 
As all reservoirs are of the ordinary type, 
the approximation in Eq.~(\ref{lindexbath}) does not hold for the end-chain baths. 
However, as one can see from Fig.~\ref{fig5therm}, 
the same pattern for the mean excitation number displayed in Fig.~\ref{fig2therm} is found.
In Fig.~\ref{fig6curr}, we then plot the currents, given in Eq.~(\ref{currssintsys}), 
generated by the attachment to the reservoirs with temperatures $T_k$, $k = 2, ..., n-1$.  
They are all negative, as in Fig.~\ref{fig3curr}, 
and sum up to 
\begin{equation}                                                                         \label{currsssysII}
\sum_{k=2}^{n-1} \mathcal{J}^{(k)}_\star = 
- (\mathcal{J}^{(1)}_\star + \mathcal{J}^{(n)}_\star) < 0.
\end{equation}
However, the actual value of the sum of the currents depends on the number of oscillators 
since $\mathcal{J}^{(1)}_\star$ and $\mathcal{J}^{(n)}_\star$ 
depends on the length of the system. 
Again, the individual currents are the difference between the energy 
stored in each oscillator and the mean energy occupation of the respective reservoir. 
As the number of oscillators  in the chain increases, 
the current and mean energy behave very much like those in Fig.~\ref{fig4thercur1}.

At this point, one might wonder about the reason for the negativity of 
the internal currents. 
Actually, it  turns out that this is a simple consequence 
of the structure of Eq.~(\ref{currssintsys}) when Eq.~(\ref{zerocond}) 
is taken into account: 
numerical explorations shows that the oscillators attached to the highest 
temperature reservoirs will have $ \bar N_k{-}\bar N^{(k)}_\star > 0$, 
while the other oscillators will not as their occupation number will also be 
determined by their own lower-temperature reservoir
and the contributions coming from the higher-temperature ones.

\begin{figure}[htbp!]                                                      
\includegraphics[width=8cm]{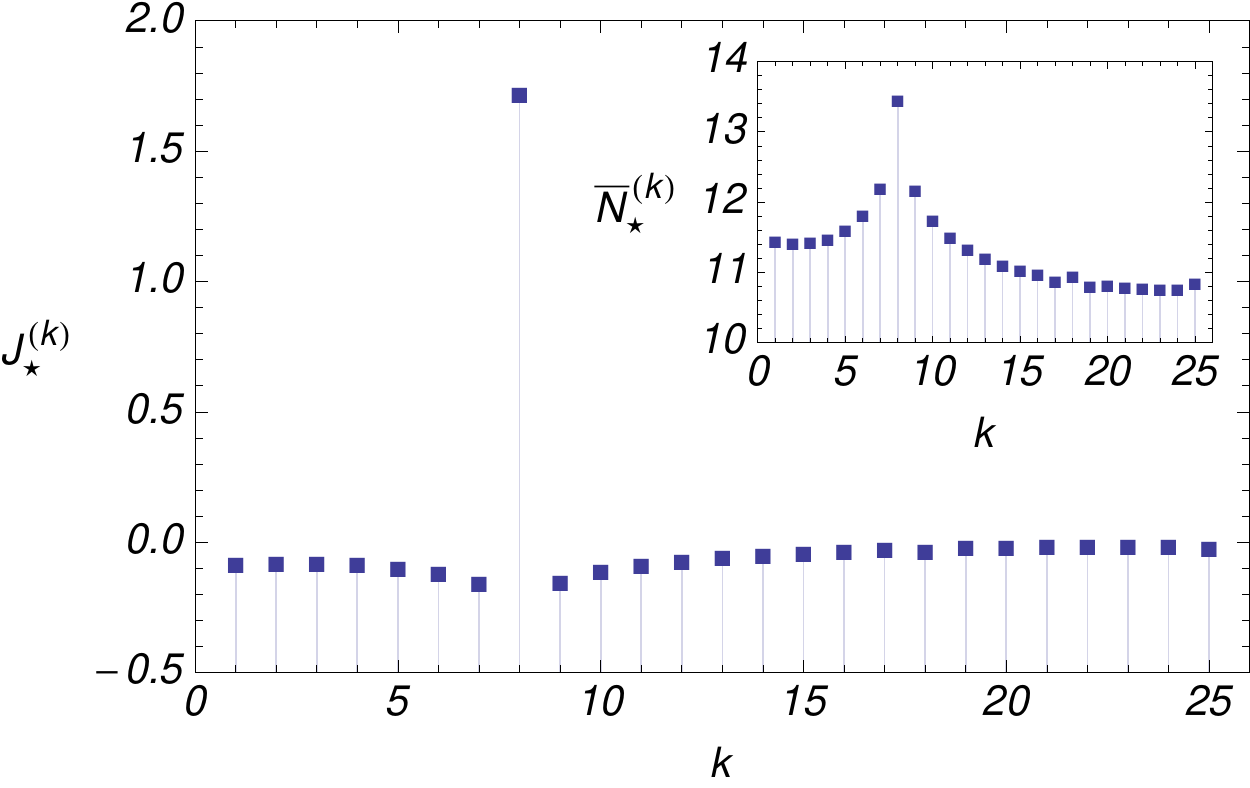} 
\caption{ 
(Color online) Currents across a chain of $ n =  25 $ oscillators with  
$\bar N_{8}  =  30$ and $\bar N_{k} = 10, \forall k \ne 8$. 
The remaining parameters are the same as in Fig.~\ref{fig2therm}. 
Inset: Mean occupation number of the oscillators in the chain.
}                                                                                        \label{fig7curr}                            
\end{figure}

In Fig.~\ref{fig7curr}, we plot the results valid for a different configuration, 
where one internal reservoir has the highest temperature. 
Notwithstanding the differences with respect to the patterns shown 
in Figs.~\ref{fig5therm} and \ref{fig6curr}, the currents and energy of this 
configuration follow the same chain-length dependence discussed above. 
%
 
\subsection{ Case II: All-diffusive dynamics } \label{caII}                 
Let us consider now a chain of oscillators connected only to the 
all-diffusive reservoirs $\rm A$ and $\rm B$. 
When $\zeta_k=0$ for $k=1,\dots,n$ the resulting dynamics is not stable ({\it cf.} Appendix). 
%
%
%
%
However, the evolution of the system can be deduced from Eq.~(\ref{cmsyst}) 
by taking $\zeta \to 0$ to give 
\begin{equation}                                                                         \label{cmsyst2}
\begin{aligned}
\!\!\!\!\!\!\!\! {\bf V}(t)  & =   {\rm e}^{\mathsf J {\bf H} t} \, {\bf V}\!_0  \, 
                          { \rm e}^{-{\bf H}\mathsf J t } \\ 
&+ {\bf O}  \! \oplus  \!  {\bf O} 
          \left( \!\! \begin{array}{rr}
                      {\bf O} {\pmb D} {\bf O} \circ {{\rm Re} \bf L}_0 & 
                    - {\bf O} {\pmb D} {\bf O} \circ {{\rm Im} \bf L}_0   \\                         
                      {\bf O} {\pmb D} {\bf O} \circ {{\rm Im} \bf L}_0 & 
                      {\bf O} {\pmb D} {\bf O} \circ {{\rm Re} \bf L}_0    
                 \end{array}   \!\!                                          \right) 
{\bf O} \! \oplus \! {\bf O}, 
\end{aligned}
\end{equation}
with $({{\bf L}_0})_{jk} = \lim_{\zeta \to 0} {\bf L}_{jk} 
= i({\rm e}^{-i (\nu_j-\nu_k) t } - 1)/( {\nu_j -\nu_k  }) $.
The diagonal elements of ${{\bf L}_0}$ are obtained taking the limit 
$\nu_j \to \nu_k$ and are given by $({{\bf L}_0})_{kk} = t, \forall k$. 
Note that the ordering of the two limits above does not commute.
As in the case of Eq.(\ref{cmsssys}), the matrix elements $({{{\rm Im} \bf L}_0})_{kk} = 0, \forall k$.
Consequently, the reduced CM of each oscillator is diagonal meaning that 
it is a thermal state for any time instant. 

\begin{figure}[t]                                                                    
\includegraphics[width=8.5cm]{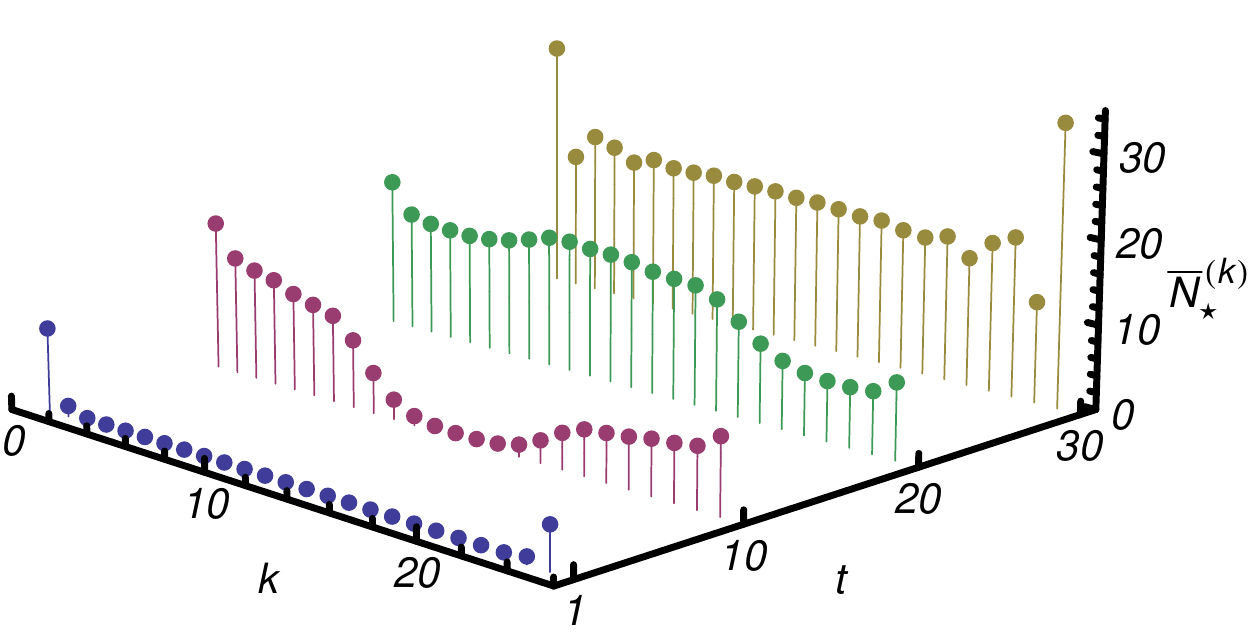}                                                  
\caption{
(Color online) Distribution of mean occupation numbers for the 
elements of a chain of $n = 25$ oscillators attached only 
to two end-chain diffusive reservoirs. 
We sample the dynamics at the instants of time $t = 1,10,20,30$. 
The chain is initially prepared in its vacuum state. 
We have taken 
$\zeta_{\rm A, B} = 1/10$ and 
$\bar N_{\rm A} = 2\bar N_{\rm B} = 100$.
As for the other parameters of the system, we have  
$\Omega/\omega = 1/2$ and $\hbar = 1$.}                                                  \label{fig8therm}
\end{figure}

In Fig.~\ref{fig8therm}, we show the mean occupation number of each oscillator, 
$\bar N^{(k)} = {\bf V}(t)_{kk}/\hbar - 1/2 $, 
at some instants of time for the evolution in Eq.~(\ref{cmsyst2}) and an 
initial vacuum state. 
The process of excitation of the elements of the chain starts from 
its ends to then progressively move towards its center. 
The mean occupation number of the oscillators increases on average linearly in time, 
{\it i.e.}, they oscillate by the effect of the orthogonal matrices $\bf O$ around the 
linear rate given by $[{{\bf L}_0}]_{kk}$ [{\it cf.} Eq.~(\ref{cmsyst2})]. 

{As an interesting remark, we observe that the distribution corresponding 
to the case of $t=20$ displays oscillators having occupation numbers larger than 
those of the oscillators in touch with the all-diffusive reservoirs. 
This is an effect of the competition between linear time increase of $\bar{N}_k$ and the 
oscillatory behavior induced by the actual absence of a steady state. }  

The total current for this system is obtained as 
\begin{equation}                                                                         \label{currsys3}
{\mathcal J}_0 := \lim_{\zeta \to 0} \mathcal J = \omega 
\left[ \zeta_{\rm A} \bar N_{\rm A}  + \zeta_{\rm B} \bar N_{\rm B} \right], 
\end{equation}
and is thus a constant and that helps us in determining the mean energy of the system
\begin{equation}                                                                         \label{meansHtsys2}
\lim_{\zeta \to 0} \langle \hat H \rangle_t = 
\tfrac{1}{2} {\rm Tr} (\mathbf H  {\bf V}_{\!0}) +  {\mathcal J}_0 \, t. 
\end{equation}
The indefinite growth of energy supplied by the current 
can be seen as a signature of instability of the system. 
In this context, the relative temperature of the two reservoirs is irrelevant since both 
contributes additively for the energy [Eq.~(\ref{meansHtsys2})] and for 
the current [Eq.~(\ref{currsys3})].  

\subsection{Case III: Balanced competition of environmental effects} \label{caIII}  
A standard thermal reservoir, 
as those distributed across the chain in Fig.~\ref{fig1chain}, 
exchanges energy with the system in two distinct ways: diffusion, 
described by the matrix $\bf D$ in Eq.~(\ref{dynmat}), which is responsible 
of the enhancement of energy; and dissipation, 
described by $\bf \Gamma$, which extracts energy from the system.  
The balanced competition of these two effects drives the system to equilibrium 
at the steady state. 

So far, the ordinary reservoirs and the all-diffusive have not been treated on equal footing, and 
it will be interesting to understand the behavior of a chain when a balanced competition
of environmental mechanisms is considered. 
To this end, let us consider the system depicted in Fig.~\ref{fig1chain} with 
$\zeta_m = 0$ for $1 < m < n$, {\it i.e.}, while the reservoirs of the bulk chain are detached, 
those at the end of it (having temperatures $T_{1}$, $T_{n}$, $T_{\rm A}$, and $T_{\rm B}$) 
are still operative. 
In this case, the matrices (\ref{dynmat}) are given by 
\begin{equation} \label{dynsys2}
{ \bf \Gamma }  =  {\pmb G} \oplus {\pmb G}   
                                + {\mathsf J} ({\pmb H} \!\oplus\! {\pmb H}),       \,\,\, 
{\bf D} = {\pmb D} \! \oplus \! {\pmb D}, 
\end{equation}
with ${\pmb G}= -\tfrac{1}{2}{\rm Diag}(\zeta_1,0,...,0,\zeta_n)$ and 
\[
{\pmb D}=  \hbar \zeta \, 
{\rm Diag}( \bar{N}_{1} + \tfrac{1}{2} + \tfrac{\zeta_{\rm A}}{\zeta}\bar{N}_{\rm A},
0,...,0,\bar{N}_{n} + \tfrac{1}{2} + \tfrac{\zeta_{\rm B}}{\zeta}\bar{N}_{\rm B}). 
\]
The stability of this system is independent of the diffusive reservoirs
since they did not contribute to ${ \bf \Gamma }$ in Eq.~(\ref{dynsys2}).
Furthermore, the action of the all-diffusive reservoirs is only to enhance 
the mean occupation number of the standard ones, that is, the solution to this problem 
is equivalent to take a chain with only two end-system standard reservoirs with mean 
occupation numbers 
$\tilde N_{1} = \bar N_{1} + ({\zeta_{\rm A}}/{\zeta})\bar{N}_{\rm A}$ and  
$\tilde N_{2} = \bar N_{2} + ({\zeta_{\rm B}}/{\zeta})\bar{N}_{\rm B}$.
The methods employed in Ref.~\cite{assadian}, 
which addressed the problem embodied by Eq.~(\ref{dynsys2}) 
with $\zeta_{\rm A} =\zeta_{\rm B} = 0 $, will be useful to find a solution to this case. 
At the steady state attained by Eq.~(\ref{dynsys2}),
the sum of all currents is null (as expected) while the mean occupation number for each 
oscillator is given by the expressions
%
%
\begin{equation}                                                                           \label{ocn}
\begin{aligned}
\bar N^{(1)}_\star &=  \tfrac{1}{2}(\tilde N_1 + \tilde N_n) + 
\frac{ \zeta^2 (\tilde N_1 - \tilde N_n) }{8\Omega^2 + 2\zeta^2 },\\ 
\bar N^{(k)}_\star &=  
\tfrac{1}{2}(\tilde N_1 + \tilde N_n)~~(1 < k < n), \\  
\bar N^{(n)}_\star &=  \tfrac{1}{2}(\tilde N_1 + \tilde N_n) - 
\frac{ \zeta^2 (\tilde N_1 - \tilde N_n) }{8\Omega^2 + 2\zeta^2 }. 
\end{aligned}
\end{equation}
All the internal oscillators have the same occupation number, as in Ref.~\cite{assadian}.  
The mean energy at the steady state can 
be evaluated from Eq.~(\ref{meansHtsys})
\begin{equation}                                                                          \label{sshsys3}
\langle \hat H \rangle_\star = 
\hbar \omega\sum_{k = 1}^n ( \bar N^{(k)}_\star + \tfrac{1}{2}) = 
n\hbar\omega\left(\frac{\tilde N_1 + \tilde N_n}{2} + \frac{1}{2}\right).
\end{equation}
{The currents through the two oscillators are given by Eq.~(\ref{currdissys}), {\it i.e.}, 
$\mathcal J^{(k)}_\star =   
-\hbar\omega \zeta (\bar N^{(k)}_\star - \tilde N_k)$ } with $k = 1,n$ 
and are equal to 
\begin{equation}                                                                         \label{sscur2}
\mathcal J^{(1)}_\star = -  \mathcal J^{(n)}_\star =   
\frac{ 2 \hbar \omega \Omega^2 \zeta (\tilde{N}_1 - \tilde{N}_n ) }
     { 4\Omega^2 + \zeta^2}.
\end{equation}
This perfect balance implies that the currents at the stationary state 
for the two standard reservoirs are constrained 
to sum up $\mathcal J^{(\rm A)} + \mathcal J^{(\rm B)}$ 
which is the sum of the currents due to the all-diffusive ones. 
This is analogous to what we have witnessed in Sec.~\ref{TSAID}. 
Since $\tilde{N}_{k} = [\exp(\hbar \beta_k \omega) - 1]^{-1}  $, 
where $\beta_k$ is the inverse temperature, one can see that,  
in the classical limit $\hbar \to 0$, the current [Eq.~(\ref{sscur2})] 
and the temperature of each reservoir in the bulk [Eq.~(\ref{ocn})] behave as in 
the classical case~\cite{rieder}. 
That means that the currents are proportional to the temperature difference and 
the bulk oscillators thermalize at the mean value temperature of the reservoirs.

%
\subsection{Case IV: Dephasing dynamics} \label{caIV}         
In Ref.~\cite{assadian}, Assadian {\it et al.} considered a chain of oscillators 
attached to two standard thermal reservoirs at the ends, which is the same
configuration described in Sec.~\ref{caIII} but with $\zeta_{\rm A} =\zeta_{\rm B} = 0 $. 
Under these restrictions, the results in Eq.~(\ref{sshsys3}) and (\ref{sscur2}) 
remain valid and can be extracted from their work. 

\begin{figure}[htbp!]                                                      
\includegraphics[width=8cm]{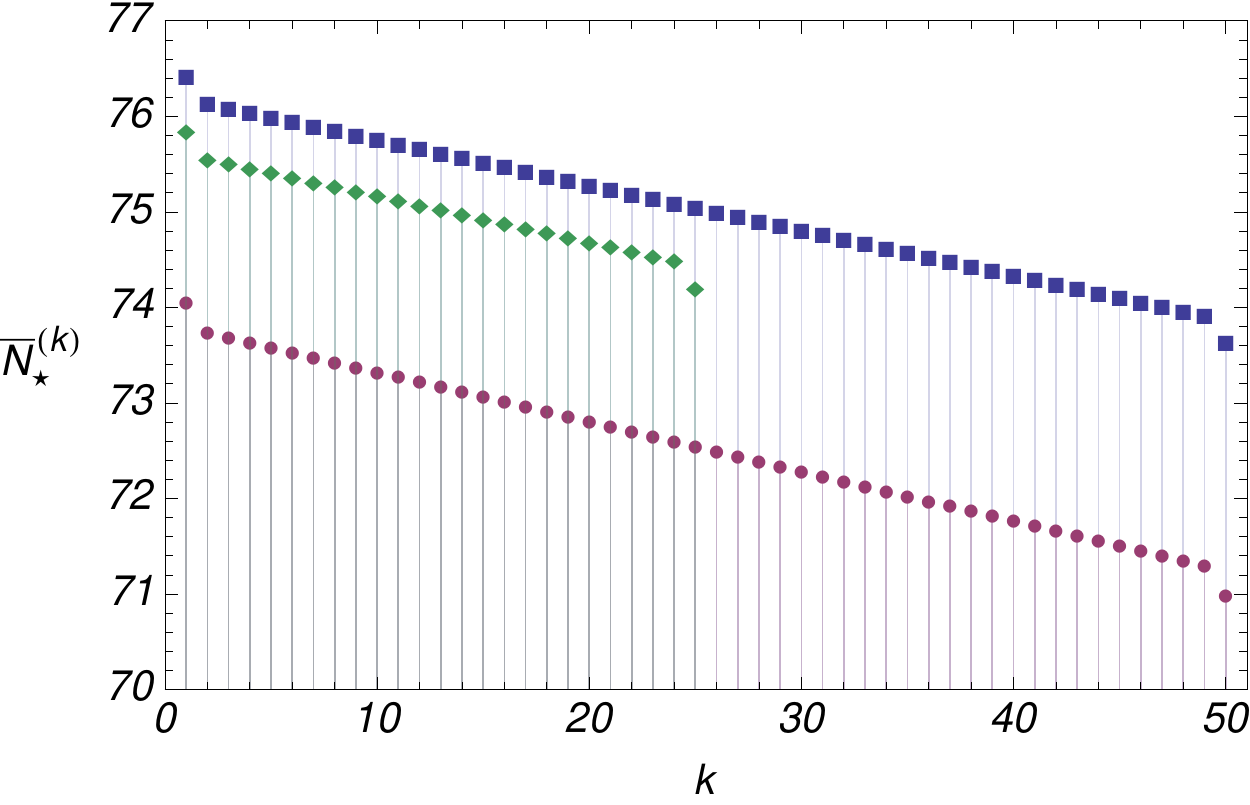} 
\caption{ 
(Color online) Distribution of mean occupation numbers for the 
elements of a chain connected to two ordinary end-chain reservoirs 
and $n$ dephasing reservoirs. 
Diamonds (green): chain with $ n =  25 $ oscillators and  bath mean 
occupation numbers $\bar N_{1}  =  2 \bar N_{n} = 100$. 
Squares (blue): chain with $ n =  50 $  oscillators and  
$\bar N_{1}  =  2 \bar N_{n} = 100$. Circles (violet): $ n =  50 $ 
and $\bar N_{1}  =  4\bar N_{n}/3 = 100$. 
All other parameters are as in Fig.~\ref{fig2therm}.}                                \label{fig9therm} 
\end{figure}

Besides this example, they consider also the presence of $n$ purely dephasing 
reservoirs, each one attached to each oscillator of the chain. 
The contribution of the dephasing mechanisms to the dynamics is modelled adding the 
following Lindblad operators to the master equation 
regulating the dynamics of the system
\begin{equation}                                                                        \label{linddeph}
\hat L_k =  \hbar \sqrt{\gamma} \, \hat a_k^\dagger \hat a_k, \,\,\, k = 1,...,n. 
\end{equation}
The special form of these reservoirs is such that they do not introduce 
new currents in the system. This can be verified by calculating the 
individual currents to find that $\mathcal J_k = 0, \forall k$. 
On the other hand, their presence drastically changes the behavior 
of the mean occupation value of each oscillator . 
As it can be seen from Eq.~(\ref{meansssys}), these are now given by \cite{assadian}
\begin{equation}
\begin{aligned}                                                                         
\bar N^{(1)}_\star &=  \tfrac{1}{2}(\bar N_1 + \bar N_n) + 
\frac{ [\zeta^2 + (n-1)\gamma \zeta] (\bar N_1 - \bar N_n) }
     { 8\Omega^2 + 2\zeta^2 + 2(n-1)\gamma \zeta           },                        \\ 
\bar N^{(k)}_\star &=  \tfrac{1}{2}(\bar N_1 + \bar N_n) + 
\frac{ (n-2k+1) \gamma \zeta (\bar N_1 - \bar N_n) }
     { 8\Omega^2 + 2\zeta^2 + 2(n-1)\gamma \zeta   },                                \\ 
\bar N^{(n)}_\star &=  \tfrac{1}{2}(\bar N_1 + \bar N_n) - 
\frac{ [\zeta^2 + (n-1)\gamma \zeta] (\bar N_1 - \bar N_n) }
     { 8\Omega^2 + 2\zeta^2 + 2(n-1)\gamma \zeta           }                         
\end{aligned}
\end{equation}
for $1 < k < n-1$. 
Note that if the temperature of standard end-chain reservoirs are equal, 
$\bar N_1 = \bar N_n = \bar N$, all the oscillators thermalize with the 
standard reservoirs having the same mean occupation number $\bar N$. 
The behavior of the mean occupation number is plotted in Fig.~\ref{fig9therm}. 

As already commented, the dephasing reservoirs 
do not contribute to the currents. 
At the stationary state, we have
\begin{equation}                                                                         \label{currsssysIII}
\mathcal{J}^{(n)}_\star = - \mathcal{J}^{(1)}_\star = 
\frac{ 2 \hbar\omega \Omega^2 \zeta (\bar{N}_1 - \bar{N}_n ) }
     { 4\Omega^2 + \zeta^2 + (n-1)\gamma \zeta}.  
\end{equation}
This result is remarkable, as it shows that for 
$ 4\Omega^2 + \zeta^2 \ll n \gamma \zeta$, which is trivially satisfied for 
a large enough chain, a Fourier-like dependence on the size of the system is recovered~\cite{assadian}. 
The classical version of the same problem has been studied in Ref.~\cite{landi}.

We now describe the modifications induced by the dephasing reservoirs
when they are attached, one by one, to the system.
The Lindblad operator in Eq.~(\ref{linddeph}) can be rewritten as (see also the Appendix)
\begin{equation}                                                                        
\hat L_m =  \tfrac{1}{2} \hat x \cdot { \Delta}_m \hat x + 
\lambda_m  \cdot \mathsf J \hat x + \mu_m ,  
\end{equation}
with the $2n \times 2n $ real matrix
\begin{equation}
[ {\bf \Delta}_m ]_{jk} = 
\sqrt{\gamma_m} (\delta_{jm} \delta_{mk} +\delta_{j+n\, m} \delta_{m\,k+n}).
\end{equation}
%
%

\begin{figure}[htbp!]                                                      
\includegraphics[width=8cm]{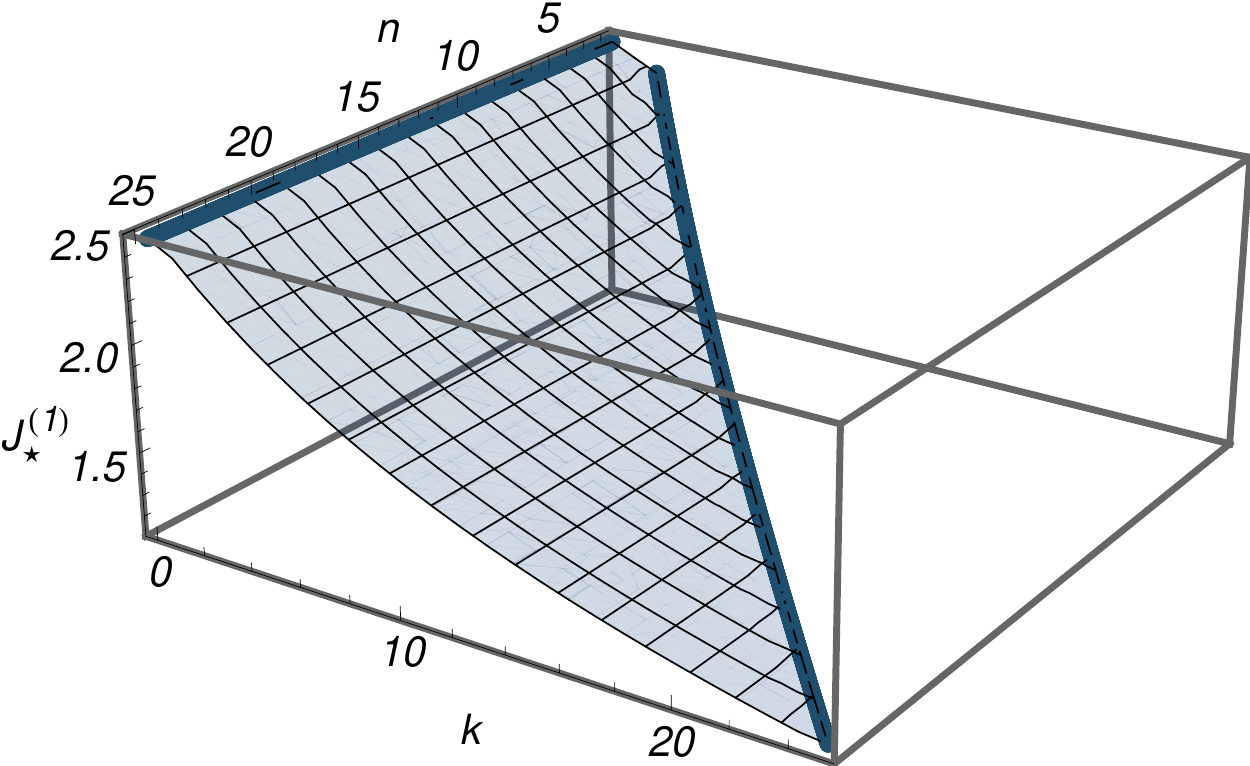} 
\caption{ 
(Color online) Current across the chain with two ordinary end-chain 
reservoirs, plotted as a function of the number of oscillators 
$n$ and the number of dephasing reservoirs $k$. 
We highlight the curves describing the extreme cases $k = 0 $ 
and $k = n$, which are the functions reported, respectively, in
Eq.~(\ref{sscur2}) with 
$ \tilde N_{1}  = 2 \tilde N_{2}  = 100 $ 
and in Eq.~(\ref{currsssysIII})  with 
$ \bar N_{1}  = 2 \bar N_{2}  = 100 $.  
The dephasing coupling is $\gamma/\omega  = 0.5$ while the remaining 
parameters are the same as in Fig.~\ref{fig2therm}. }                                                 \label{fig10curr}                            
\end{figure}

Without dephasing reservoirs ($\gamma_m = 0, \forall m$), the currents in the system 
are the ones described by (\ref{sscur2}) with the substitutions 
$\tilde N_{1} \to \bar N_{1}$ and $\tilde N_{n} \to \bar N_{n}$.  
Following the prescriptions in the Appendix, 
we solve numerically the system with 
$\gamma_1 = \gamma$ and $ \gamma_k = 0 \, \forall k \ne 1$, $n \le k\le 1$ 
and calculate the current as function of the number of oscillators $n$. 
Adding progressively more dephasing reservoirs until $\gamma_m = \gamma, \forall m$  
[in this situation the current is given by Eq.~(\ref{currsssysIII})]
and calculating the current allows us to show in Fig.~\ref{fig10curr} the smooth transition 
from the situation described by Eq.~(\ref{sscur2}) to that associated to Eq.~(\ref{currsssysIII}). 

\subsection{Case V: Disorder Effect} \label{caV}           
Classically, size-dependent currents in chains of oscillators arise under the presence of 
anharmonicity or disorder~\cite{dhar}, which can be realized in various ways. 
One can, for example, introduce different frequencies and/or couplings across the chain.
For the sake of definiteness, we consider the system of Fig.~{\ref{fig1chain}}, 
now ruled by the Hamiltonian 
\begin{equation}                                                                         \label{randh}
\hat H =  \hbar\omega \sum_{j = 1}^n \hat{a}_j^\dag \hat{a}_j + 
           2 \hbar \sum_{j = 1}^{n - 1} \Omega_j 
            (\hat{a}_j^\dag \hat{a}_{j+1} + \hat{a}_{j+1}^\dag \hat{a}_j). 
\end{equation}

The structure of Eq.~(\ref{cmsssys}) remains the same with 
$\bf O$ being replaced by the matrix that diagonalizes the adjacency matrix    
\begin{equation}                                                                         \label{randh2}
{\pmb H}_{\! j k} = \omega \, \delta_{j k} + 
                    \Omega_j \, (\delta_{j \, k+1} + \delta_{j \,k-1} ) 
\end{equation}
and $\nu_k$ (which appears in the definition for ${\bf L}_\star$)  
being its eigenvalues and both can be calculated numerically
for a given set of couplings $ \{\Omega_j\}_{1 \le j \le n} $.   
To introduce disorder, we arbitrary choose a set of coupling constants 
$\Omega_j$ with magnitudes similar to what has been considered so far.

\begin{figure}[t]                                                                                                                         
\includegraphics[width=8cm]{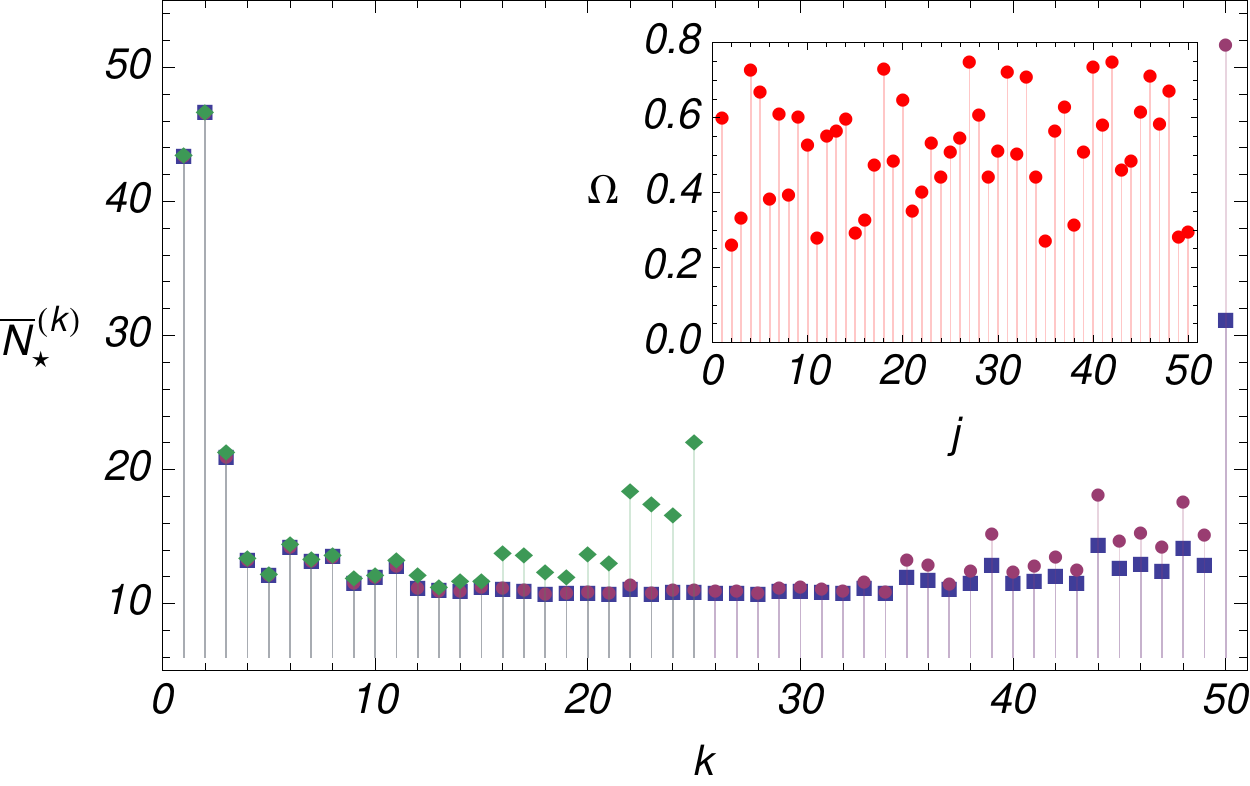}                                                 
\caption{ 
(Color online) Distribution of mean occupation numbers for the elements 
of a chain with distinct coupling constants. 
Inset: Set of coupling constants used in the simulation. 
The remaining parameters are as Fig.~\ref{fig2therm}.}                    \label{fig11therm}          
\end{figure}

\begin{figure}[t]                                                      
\includegraphics[width=8cm]{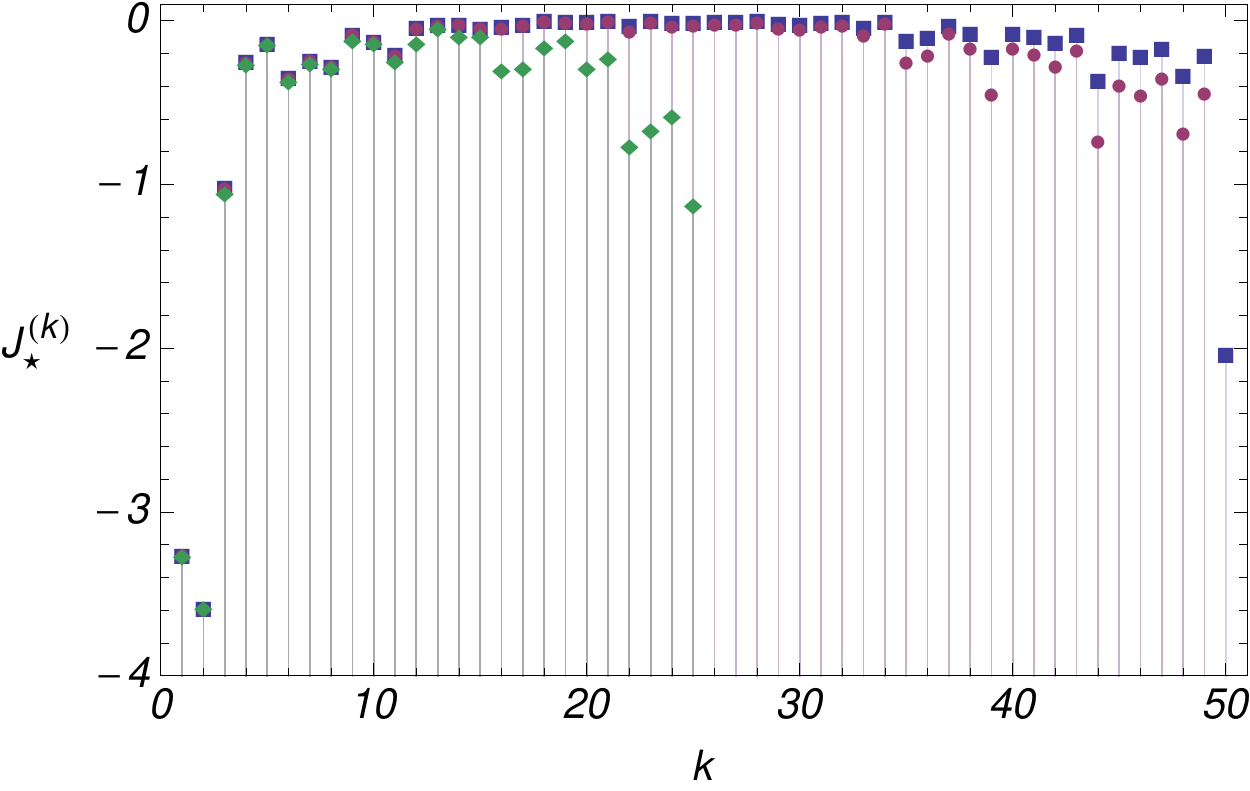} 
\caption{ 
(Color online) Currents across the chain for the same situation depicted in  
Fig.~\ref{fig11therm}.}                                                                  \label{fig12curr}                            
\end{figure}

In Figs.~\ref{fig11therm} and \ref{fig12curr}, respectively, 
we plot the mean occupation number and the current for a set of distinct 
coupling constants.  
Structurally speaking, the behavior of the system does not 
change with the introduction 
of disorder. 
This can be seen when comparing these figures, respectively, 
with Fig.~\ref{fig2therm} and Fig.~\ref{fig3curr}. Observe the similarities on
the profile of the curves, 
the first, and last oscillators and, mainly, the behavior of the bulk. 
Furthermore, for the quantities shown in Figs.~\ref{fig11therm} and \ref{fig12curr}, 
we also have analyzed different sets of inhomogeneous constant couplings $\{\Omega_j\}$. 
In the units of the paper, each $\Omega_j$ is bounded by $(0,1]$ 
in our simulations in order to keep a fair comparison of the results. 
From these simulations, results obtained using different sets differ, 
but the general trends shown in those plots are preserved,
especially the behavior of the bulk oscillators. 


\begin{figure}[htbp!]        
\includegraphics[width=8.5cm]{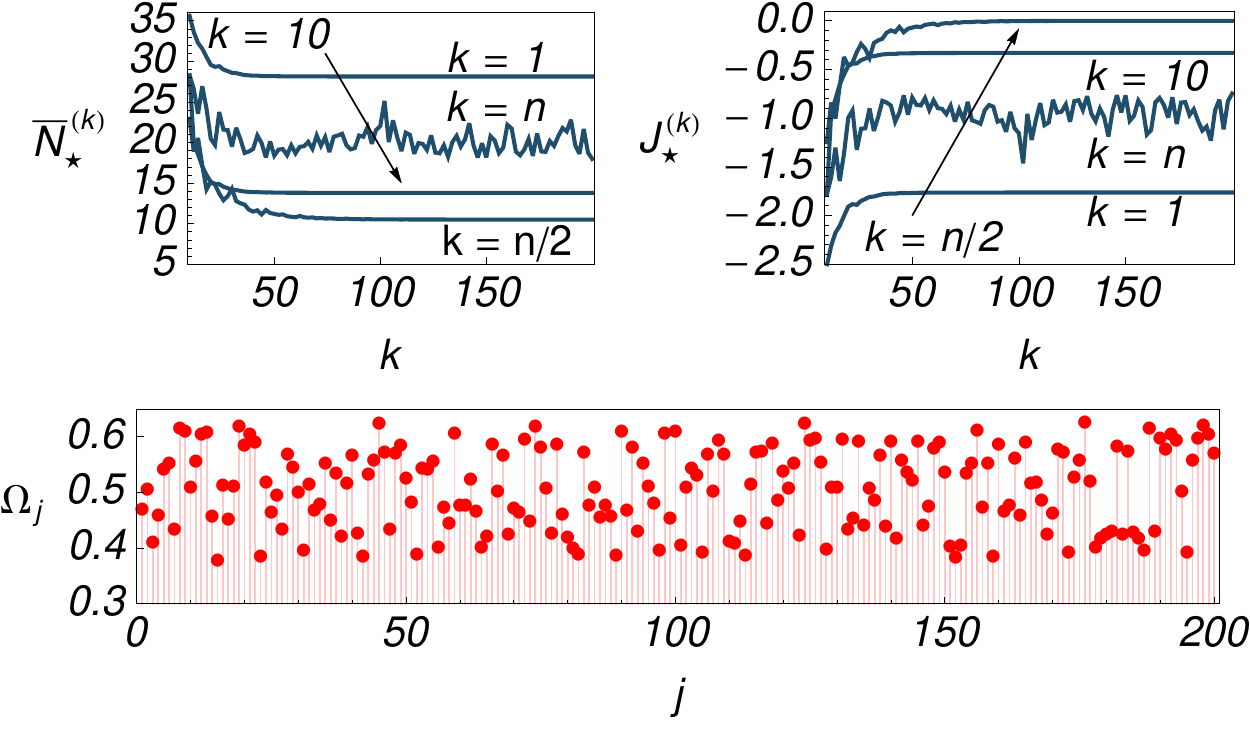} 
\caption{ 
(Color online) 
Mean value of energy (left) and currents (right) 
for fixed position oscillators 
[the first $( k = 1 )$ and the 10$^\text{th}$ $(k = 10)$],  
the midpoint (bulk) $ k = n/2 $, and the last oscillator $k = n$ 
for a chain with distinct constant couplings.  
Bottom panel: Set of coupling constants used in the simulation. 
We consider $ \bar N_{\rm A}  =  2 \bar N_{\rm B} = 10 \bar N_k = 100 $. 
The remaining parameters are the same as in Fig.~\ref{fig2therm}.}                       \label{fig13thercur2}                            
\end{figure}

It is also interesting to see what happens with the individual currents
and occupation numbers near the thermodynamical limit. 
These are plotted in Fig.~\ref{fig13thercur2}. 
As before, currents and mean occupation numbers for the oscillators in the bulk become 
independent on the length for long-enough chains. 
However, it is interesting to remark that the last oscillator is strongly influenced by 
the disorder while the behavior of the remaining oscillators 
is essentially the same as in the case of Fig.~\ref{fig4thercur1}. 

\subsection{Case VI: Spring-mass coupling} \label{caVI}       
Until now, we have worked in the regime of the rotating-wave approximation (RWA), 
which enables the explicit analytical form of the CM of the chain [{\it cf.}~Eq.(\ref{cmsyst})]. 
This contrasts significantly with any classical approach to the transmission of 
heat across a harmonic chain, which is in general performed 
assuming the standard spring-mass coupling (SMC)~\cite{rieder,dhar,landi}. 
In this subsection we will thus briefly address the case of an SMC-like coupling 
to make a more faithful comparison with the classical case.

\begin{figure}[htbp!]                                                      
\includegraphics[width=8cm]{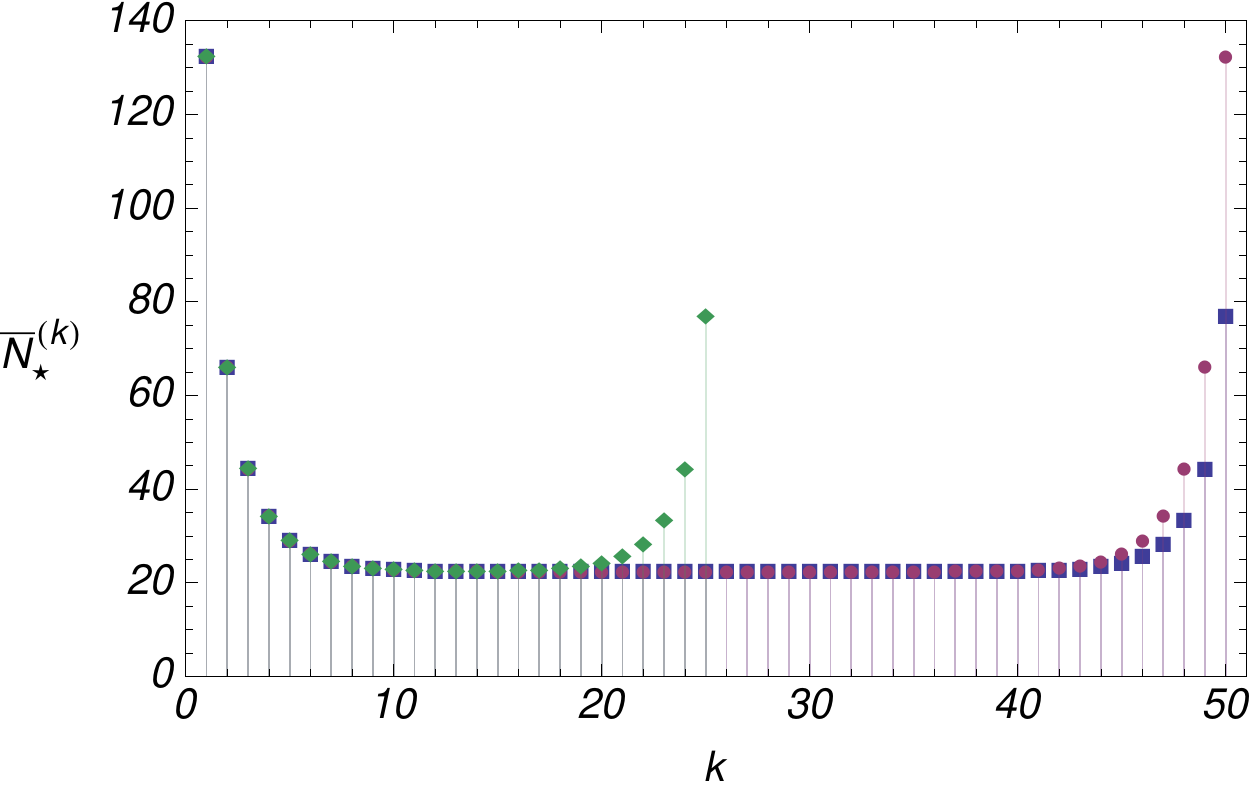} 
\caption{ 
(Color online) Distribution of mean occupation numbers for the 
elements of a chain of two lengths for the SMC Hamiltonian. 
Diamonds (green): chain of $ n =  25 $ with 
$\bar N_{\rm A}  =  2 \bar N_{\rm B} = 10 \bar N_k = 100$;  
Squares (blue): chain of $ n =  50 $ with  
$\bar N_{\rm A}  =  2 \bar N_{\rm B} = 10 \bar N_k = 100$; 
Circles (violet): $ n =  50 $  with  
$\bar N_{\rm A} =   \bar N_{\rm B} = 10 \bar N_k = 100$.  
The remaining parameters are $\kappa/\omega = 1/2$, 
$\zeta/\omega = \zeta_{\rm A}/\omega = \zeta_{\rm B}/\omega = 1/10$, 
and $\hbar = 1$.}                                                                          \label{fig14therm} 
\end{figure}

\begin{figure}[htbp!]                                                      
\includegraphics[width=8cm]{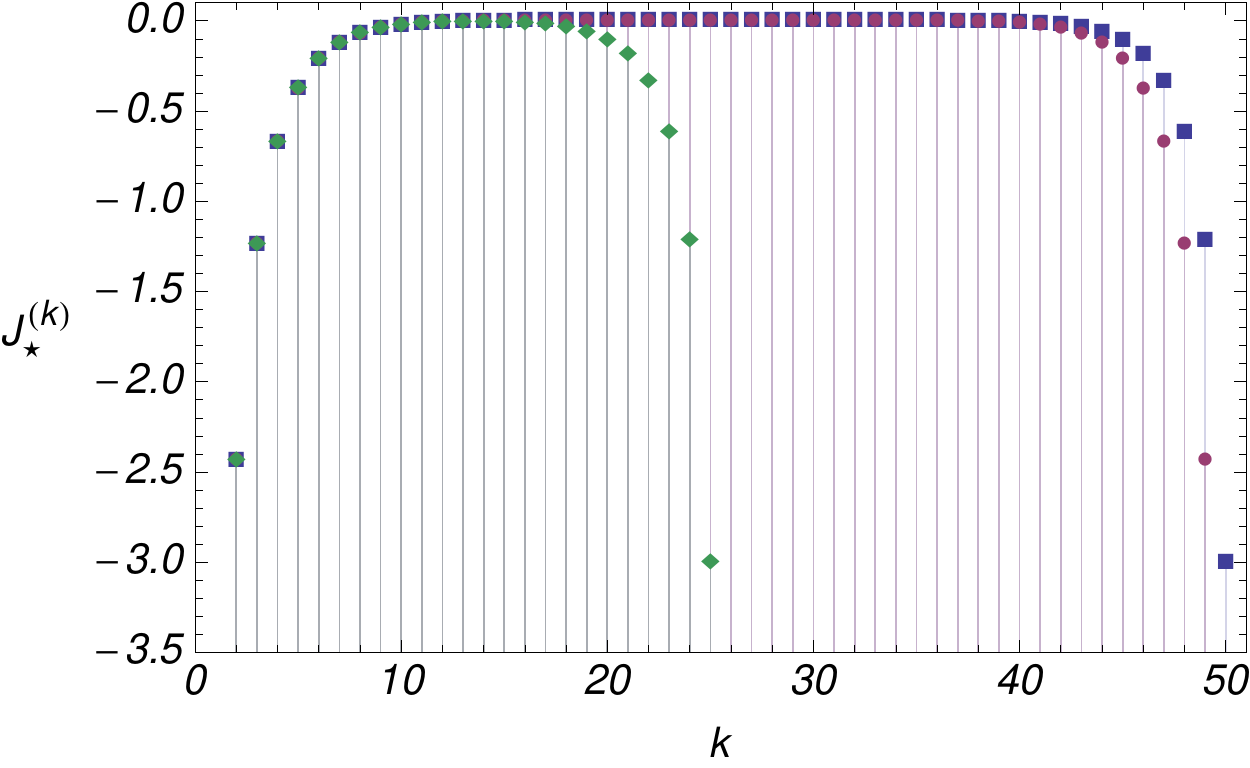} 
\caption{ 
(Color online) Currents across the chain for the SMC Hamiltonian. 
We used the same parameters as in  Fig.~\ref{fig14therm}. 
}                                                                                        \label{fig15curr}                            
\end{figure}

The Hamiltonian of a chain of $n$ oscillators coupled by the standard SMC coupling 
has the adjacency matrix 
$\mathbf H' = {\pmb H}' \oplus \omega \mathsf I_n$, where 
\begin{equation}                                                                         \label{hamsys3}
{\pmb H}'_{\! j k}  = \, ( \omega + \kappa ) \, \delta_{j k} 
                    -  \tfrac{\kappa}{2} ( \delta_{j\, k\pm1} 
                   + \delta_{j 1} \delta_{1 k} + 
                                           \delta_{j n} \delta_{n k}).
\end{equation}
As this matrix is almost of the Toeplitz form, a procedure similar to the one used in 
the Appendix can be used to find 
%
the covariance matrix of the system, which is given by
\begin{equation}
{\bf V} = {\bf O }_{{\bf \Gamma}'}^{-1} 
          \left[ { \bf O }_{{\bf \Gamma}'} 
                 { \bf D } 
                 { \bf O }_{{\bf \Gamma}'}^\top \circ {{\bf L}'_\star}                                                             
         \right] {\bf O }_{{\bf \Gamma}'}^{-\top}.
\end{equation}
Here $[{\bf L}'_\star ]_{jk} = -1/(\nu_j' - \nu_k' ) $  with
\begin{equation}
\nu'_k = \sqrt{ \omega ( \omega + \kappa ) - 
         \omega\kappa \cos\left[ (m-1)\pi/n\right]},
\end{equation}
and  
${\bf O }_{{\bf \Gamma}'}$ defined by  
\begin{equation}
\!\!{\bf O }_{{\bf \Gamma}'} {\bf \Gamma}'{\bf O }_{{\bf \Gamma}'}^{-1}  = 
-\frac{\zeta}{2} \mathsf I_{2n} + 
{\rm Diag}(i\nu'_1,...,i\nu'_n, -i\nu'_1,...,-i\nu'_n   ).
\end{equation}
%
We can now calculate the mean occupation number and the current for each oscillator,
whose behavior is shown in Fig.~\ref{fig14therm} and Fig~\ref{fig15curr}, respectively.
%
%

As one can see, the results for the SMC case are structurally similar 
to those found in the RWA. 
Such similarities extend also to the situations where either the diffusive baths 
are not considered or the distribution of coupling constants across the chain is not uniform. 

\section{Conclusions}\label{conc}
We have investigated heat transport in quantum harmonic chains connected to different 
types of heat baths. We have obtained the exact expression for the currents across the 
system highlighting the crucial role played by the properties of the environment. 
Such detailed analysis was instrumental to the study of a few paradigmatic configurations. 
In particular, just like in Ref.~\cite{assadian}, 
we have found that the Fourier law is not predicted by the models considered 
here unless dephasing is included, destroying the ballistic behavior. 
This is akin to a substrate external potential in the classical version of 
the problem addressed here.  In our findings, Fourier law is still not observed 
in the presence of disorder as long as the coupling with the bath is weak, allowing a 
description using a Markovian master equations, as addressed in this manuscript.
Our results are consistent with those known for unidimensional classical harmonic chains 
and shed new light on the interplay between transport properties of spatially extended 
quantum media and the nature of the environmental systems interacting with it. 

\acknowledgments 
FN, FLS and MP are supported by the CNPq ``Ci\^{e}ncia sem Fronteiras'' 
programme through the ``Pesquisador Visitante Especial'' initiative 
(grant nr. 401265/2012-9).
MP acknowledges financial support from the UK EPSRC (EP/G004579/1). 
MP and AF are supported by 
the John Templeton Foundation (grant ID 43467), and the EU Collaborative Project TherMiQ (Grant Agreement 618074). 
AI and MP gratefully acknowledge support from  the COST Action MP1209 
``Thermodynamics in the quantum regime".
AI is supported by the Danish Natural Science Research Council.
FLS is a member of the Brazilian National Institute of Science and Technology of 
Quantum Information (INCT-IQ) and acknowledges partial support from CNPq 
(grant nr. 308948/2011-4).
%

\renewcommand{\theequation}{A-\arabic{equation}} 
\setcounter{equation}{0} 
\section*{Appendix}                          
In this Appendix we provide additional details on the mathematical approach to the problems 
addressed in the main body of the paper. 

\subsection*{Currents and energy for a quadratic system}\label{carfls}

Here we address the derivation of the expressions for the 
currents and mean energy for a system evolving according to Eq.~(\ref{lindblad}) 
when a quadratic Hamiltonian as the one in Eq.~(\ref{ham-lind}) 
and quadratic Lindblad operators are considered. 
Specifically, we assume the form
\begin{equation}                                                                         \label{Alind}
\hat L_m =  \tfrac{1}{2} \hat x \cdot { \Delta}_m \hat x + 
\lambda_m  \cdot \mathsf J \hat x + \mu_m ,  
\end{equation}
where $\Delta_m = \Delta_m^\top $ is a $2n \times 2n $ real matrix. 
Taking the derivative of $\langle \hat x \rangle_t$ and $\bf V$ 
defined in Eq.~(\ref{cmdef}), 
using the master equation (\ref{lindblad}), and recalling the 
commutation relation $[\hat x_j , \hat x_k ] = i \hbar \, \mathsf J_{jk}$, 
we get the dynamical equations
\begin{equation}                                                                         \label{Amvcm2}
\frac{d\langle \hat x \rangle_t}{d t}  = 
\xi - \eta + { \bf \tilde\Gamma } \langle \hat x \rangle_t,~~\frac{d \mathbf V}{d t} = 
{ \tilde{\bf \Gamma} }\mathbf V + \mathbf V {\tilde{\bf \Gamma}}^\top  +  
{\bf D}  + {\bf \Delta}_{\!\bf V},  
\end{equation}
where
\begin{equation}
{\bf \Delta}_{\!\bf V} = \hbar \sum_{ m} \mathsf J {\Delta}_m  {\bf V} \, 
       {\Delta}_m \mathsf J^\top,~~\tilde{\bf \Gamma}  = {\bf \Gamma} + 
       \tfrac{\hbar}{2} \sum_{m } ( \mathsf J \Delta_m)^2.
\end{equation}
Both $\bf \Gamma$ and $\bf D$ are defined in Eq.~(\ref{dynmat}). 
By inserting Eq.~(\ref{ham-lind}) and Eq.~(\ref{Alind}) into the 
definition of the individual ${\mathcal J}_m$'s in Eq.~(\ref{totcurr1}), 
we get 
%
%
\begin{eqnarray}                                                                         \label{Apartialcurr}
\mathcal J_m & = & \tfrac{\hbar}{2}  
{\rm Tr} \left[  
               {\bf H } {\rm Re}\left( \lambda_m \lambda_m^\dag  \right) -
               {\bf H } \mathsf J {\Delta}_m  {\bf V}
                                        {\Delta}_m \mathsf J   \right] \nonumber \\
             & + &  \tfrac{\hbar}{2} {\rm Tr} \left[ {\bf H}( \mathsf J \Delta_m)^2     
           \left( {\bf V} + 
           \langle \hat x \rangle_{t} \langle \hat x \rangle_{t}^\top 
           \right) \right]                                         \nonumber \\ 
             & - &  {\rm Tr} \left[ {\bf H}\,
                                {\rm Im} \left(\lambda_m \lambda_m^\dag \right)  \mathsf J
           \left( {\bf V} + 
           \langle \hat x \rangle_{t} \langle \hat x \rangle_{t}^\top 
           \right)\right]                                         \nonumber \\
&+&  \mathsf J\xi \cdot \left[{\rm Im} 
\left(\lambda _m \lambda_m^\dag  \right) 
\mathsf J - \hbar \mathsf J {\Delta}_m  {\bf V} \, 
{\Delta}_m \mathsf J \right] \langle \hat x \rangle_t  \nonumber \\
& + & {\rm Im} (\mu_m^\ast  \lambda{_m} ) \cdot 
(\mathsf J\xi - \mathbf H \langle \hat x \rangle_t). 
\end{eqnarray}
Summing over all Lindblad operators, see Eq.~(\ref{totcurr1}), 
the total current has the following form
\begin{eqnarray}                                                                         \label{Atotcurr2}
\!\!\!\!\!\!\!\!\!\mathcal J  = & & \,  \frac{1}{2}  
{\rm Tr} \left[ {\bf H }( \, {\bf D } + {\bf \Delta}_{\!\bf V} ) \right] 
+ {\rm Tr} \left[ {\bf H} \, { \tilde{\bf \Gamma} } 
                   \left(  {\bf V} + 
                   \langle \hat x \rangle_{t} \langle \hat x \rangle_{t}^\top 
                   \right)  
           \right] \nonumber \\
& & + \, (\xi - \eta) \cdot  \mathbf H \langle \hat x \rangle_t 
- \mathsf J\xi \cdot 
 \tilde{\bf \Gamma}\langle \hat x \rangle_t, 
\end{eqnarray}
which is zero for a possible steady state, {\it i.e.}, 
the solution of (\ref{Amvcm2}) with 
$ \partial_t{\langle \hat x \rangle_t} = \partial_t{\mathbf V} = 0$. The internal energy of the system is easily worked out as
$\langle \hat H \rangle_t = {\rm Tr}( \hat H \hat\rho )$. 
We get  
\begin{equation}                                                                         \label{AmeanHt}
\langle \hat H \rangle_t = 
\tfrac{1}{2}{\rm Tr}\left[ {\bf H}\,{\bf  V}(t)  + 
{\bf H}\,\langle \hat x \rangle_t \langle \hat x \rangle_t^\top \right] +
\xi \cdot \mathsf J \langle \hat x \rangle_t + H_0.
\end{equation}
Taking the derivative of this equation and rearranging the expressions 
one also finds Eq.~(\ref{Atotcurr2}). 

\subsection*{On the stability of the dynamical system}
Here we analyze the dynamical stability of the system discussed in Sec.~\ref{TSAID}. 

The matrix $\pmb H$ given by Eqs.~(\ref{hamsys2}) and appearing in (\ref{dynsys}) 
is tridiagonal and symmetric Toeplitz, and can thus be diagonalized by a simple 
(symmetric) orthogonal transformation~\cite{kulkarni} as
$ { \bf O \pmb H \bf O^\top } = {\rm Diag}(\nu_1,...,\nu_n )  $,
where 
\begin{equation}                                                                         \label{toep}
\begin{aligned}
{\bf O}_{kl} = \sqrt{\tfrac{2}{n+1}} \, \sin \tfrac{k l \,\pi}{ n + 1},~~\nu_m = \omega + 2 \Omega \cos\tfrac{m \,\pi}{ n + 1}.   
\end{aligned}
\end{equation}
The matrix that diagonalizes $\bf \Gamma$ in (\ref{dynsys}) can 
then be constructed as 
${\bf O}_{\bf \Gamma} =  ( {\bf  O} \! \oplus \! {\bf O } ) {\bf U }$ with
\begin{equation}                                                                         \label{matrixU}
{\bf U } = \frac{1}{\sqrt{2}} 
\left(\begin{array}{rl}
      - i \mathsf I_n & \mathsf I_n \\
        i \mathsf I_n & \mathsf I_n
      \end{array}
\right),
\end{equation}
which gives us
\begin{equation}                                                                         \label{gammaeig}
{\bf O}_{\bf \Gamma} {\bf \Gamma} {\bf O}_{\bf \Gamma}^\dagger =
-\tfrac{\zeta}{2}\mathsf I_{2n} - 
i \, {\rm Diag}(\nu_1,...,\nu_n,-\nu_1,...,- \nu_n ).
\end{equation}
As the spectrum in Eq.~(\ref{gammaeig}) has positive real part, $\bf \Gamma $ is stable. 
Moreover, in light of the fact that 
$\mathbf D$ in Eq.~(\ref{dynsys}) is a positive definite matrix, the system allows 
for a steady state, whose moments are given by (\ref{mvcmss}). 


\

\subsection*{Details on the calculation of Eq.~(\ref{cmsolt})} \label{doce18}

By integrating the expression for ${\bf I}$ by parts, we get  
\begin{equation}
i [ {\bf I}, {\pmb H} ] - \zeta {\bf I} = 
{\rm e}^{-i {\pmb H} t} {\pmb D } {\rm e}^{i {\pmb H} t} {\rm e}^{-\zeta t}  - {\pmb D}. 
\end{equation}
By diagonalizing ${\pmb H}$ with the help of Eq.~(\ref{toep}) 
and introducing the matrices $\tilde{\bf I} = {\bf O} {\bf I} {\bf O}$,  
$\tilde{\pmb D} = {\bf O} {\pmb D} {\bf O}$, and $\bf L$ given 
in Eq.~\eqref{matrixOL}, we find that 
$  \tilde{\bf I}_{jk} = \tilde{\pmb D}_{\!jk} {\bf L}_{jk}  =  
 (\tilde{\pmb D} \circ {\bf L})_{jk} $. 
Starting from this, one can straightforwardly show that  
\begin{equation}                                                                         \label{matrixI}
{\bf I} = {\bf O} 
          \left[{\bf O} {\pmb D} {\bf O} \circ {\bf L}\right] {\bf O}. 
\end{equation}
In turn, the dynamical solution in Eq.~(\ref{cmsolt}) 
can be obtained by using this result and noticing that 
\begin{equation}
{\bf U}^\dagger \,( {\bf I} \oplus {\bf I}^\ast ) \,  {\bf U} = 
\left(\begin{array}{cr}
      {\rm Re} {\bf I} & - {\rm Im} {\bf I} \\
      {\rm Im} {\bf I} & {\rm Re} {\bf I}
      \end{array}
\right). 
\end{equation}
%

\subsection*{Structural Properties of the CM}

We now analyze some structural details of the blocks 
of the CM in Eq.~(\ref{cmsssys}), which with the help of (\ref{matrixI}) becomes 
\begin{equation}                                                                         \label{cmblock}
{\bf V}_{\!\star}  = 
\frac{\hbar}{2}\mathsf I_{2n} \, + 
\begin{pmatrix}
{\rm Re} {\bf I}_\star  & - {\rm Im} {\bf I}_\star \\
{\rm Im} {\bf I}_\star &  {\rm Re} {\bf I}_\star 
\end{pmatrix} , 
\end{equation}
where ${\bf I}_\star = {\lim_{t\to \infty}} {\bf I}$ 
is a $n\times n$ Hermitian matrix. 
Actually, we want to demonstrate that ${\bf I}_\star$ satisfies   
\begin{equation}                                                                         \label{result}
\begin{aligned}                                                                         
&{{\rm Re} {\bf I}_\star}_{jk} = 0 \,\,\, \text{if} \,\,\, j + k \,\,\, \text{ is odd} , \\  
&{{\rm Im} {\bf I}_\star}_{jk} = 0 \,\,\, \text{if} \,\,\, j + k \,\,\, \text{ is even},  
\end{aligned}
\end{equation}
which also proves the assertion on Eq.~(\ref{zerocond}).  

We start by defining the function 
\begin{equation}                                                                         \label{auxdef1}
\Phi_{jk}(l,m,r) := {\bf O}_{jl} {\bf O}_{lm} {\pmb D}^{(m)} 
                    {\bf O}_{mr} {\bf O}_{rk} {\bf L_\star}_{lr} 
\end{equation}
with ${\pmb D}^{(m)} = {\pmb D}_{mm}$, 
in such a way that from Eq.~(\ref{matrixI}) 
one can write
\begin{equation}                                                                         \label{sum1}
{{\bf I}_\star}_{jk} =  \sum_{l,m,r = 1}^{n} \Phi_{jk}(l,m,r). 
\end{equation}
The desired result, Eq.~(\ref{result}), follows from proper reorganization 
of the indexes involved in the summation, as outlined bellow. 
First consider the case $n$ {\it even}. 
In this situation, we rewrite Eq.~(\ref{auxdef1}) as 
\begin{eqnarray}                                                                        \label{sum2}
\!\!\!\! {{\bf I}_\star}_{jk}  = 
\sum_{m = 1}^{n} \sum_{l,r = 1}^{n/2} 
&\left[ \,\,\,   \Phi_{jk}(l,m,r)  + \Phi_{jk}(l',m,r') \, + \right. & \nonumber \\
&\left.  \Phi_{jk}(l',m,r)  + \Phi_{jk}(l,m,r')      \,\, \right]   &                                
\end{eqnarray}
with $l' : = (n+1 - l)$ and $r' : = (n+1 - r)$. 
From the definitions in Eq.~(\ref{matrixOL}), one can show straightforwardly that 
\begin{equation}
\begin{aligned}
&{\bf O}_{jl'} = {\rm e}^{i j\pi} {\bf O}_{jl}\, , \,\,\,  
{\bf O}_{r' k} = {\rm e}^{i k\pi} {\bf O}_{rk}\, , \\
&{\bf L_\star}_{l' r'} = { \bf L^{\ast}_\star}_{l r}\, , \,\,\, 
{ \bf L_\star}_{l r'} =  { \bf L^{\ast}_\star}_{l' r}\, ,   
\end{aligned}
\end{equation}
and Eq.~(\ref{sum2}) becomes 
\begin{eqnarray}                                                                         \label{sum3}
\!\!\!\!\!\!\!\!\!\!\!\!\!\!\!\!\!\!\!\!\!  {{\bf I}_\star}_{jk}  = 
\sum_{m,l,r} 
&&\left[ \Phi_{jk}(l,m,r)  +  {\rm e}^{i\pi(j+k)}\Phi_{jk}^\ast(l,m,r) \, + \right. \nonumber \\
&&\,\,\left. \Phi_{jk}(l',m,r) +  {\rm e}^{i\pi(j-k)}\Phi^\ast_{jk}(l',m,r)      \right],  
\end{eqnarray}
where the indexes of the sum span the same set as in (\ref{sum2}).
Finally, if $j+k$ is odd (even), $|j-k|$  is odd (even), and the above 
sum is purely imaginary (real), which proves the statement (\ref{result}) 
for the case $n$ {\it even}.   

The proof for $n$ {\it odd} follows the same steps but with (\ref{sum1})
rewritten as
%
%
%
%
%
\begin{equation}                                                                         \label{sum4}
{{\bf I}_\star}_{jk}  = {{\bf I}_\star}_{jk}' + {{\bf I}_\star}_{jk}'' 
+ \sum_{m=1}^{n}  \Phi_{jk}(\tfrac{n+1}{2},m,\tfrac{n+1}{2}),                      
\end{equation}
where ${{\bf I}_\star}_{jk}'$ is the sum in (\ref{sum3}) with the index running over
the sets $1 \le l,r  \le (n-1)/2$ and $1 \le m  \le n$. Also, 
\begin{eqnarray}                                                                         \label{sum4.1}
&&{{\bf I}_\star}_{jk}'' :=                         \\
&& \sum_{m = 1}^{n} \sum_{l = 1}^{\tfrac{n-1}{2}} 
\left[ \Phi_{jk}(l,m,\tfrac{n+1}{2} )  +       
         {\rm e}^{i\pi(j+m)}\Phi_{jk}^\ast(l,m, \tfrac{n+1}{2}) \, + \right.             \nonumber \\
& & \left.  \,\,\,\,\,\,\,\,\,\,\,\,\,\,\,\,\,\,\,\,\,\,\,\,\,  
           \Phi_{jk}(\tfrac{n+1}{2},m,l) +  
           {\rm e}^{i\pi(k+m)}\Phi^\ast_{jk}(\tfrac{n+1}{2},m,l)   \,\,   \right].      \nonumber                      
\end{eqnarray}
 
Now, we should give a closer look to the terms in Eq.~(\ref{sum4}).  
From (\ref{auxdef1}) and Eq.~(\ref{matrixOL}), it is straightforward 
to show that
%
\begin{equation}                                                                         \label{term1}
\Phi_{jk}(\tfrac{n+1}{2},m,\tfrac{n+1}{2}) \propto
\sin(j\tfrac{\pi}{2})\sin(k\tfrac{\pi}{2}) 
\end{equation}
and real for all $m$. 
For $j + k$ odd, either $j$ or $k$ must be even, 
causing this function to be zero for any value of $m$. 

On the other hand, since 

\begin{eqnarray}                                                                         \label{term2}
&&\Phi_{jk}(l,m,\tfrac{n+1}{2} ) \propto 
\sin(k\tfrac{\pi}{2})\sin(m\tfrac{\pi}{2}),              \nonumber \\
&&\Phi_{jk}(\tfrac{n+1}{2},m,l) \propto   
\sin(j\tfrac{\pi}{2})\sin(m\tfrac{\pi}{2}),                          
\end{eqnarray}
one can rewrite (\ref{sum4.1}) as
\begin{eqnarray}                                                                         \label{term3}
&&{{\bf I}_\star}_{jk}'' =  
\sum_{m = 1}^{n} \sum_{l = 1}^{\tfrac{n-1}{2}}
\left[ \Phi_{jk}(l,m,\tfrac{n+1}{2} )  -       
{\rm e}^{i\pi j}\Phi_{jk}^\ast(l,m, \tfrac{n+1}{2}) \, + \right.                 \nonumber \\
&&\,\,\,\,\,\,\,\,\,\,\,\,\,\,\,\,\,\,\,\,\,
\left. \Phi_{jk}(\tfrac{n+1}{2},m,l) -  
       {\rm e}^{i\pi k}\Phi^\ast_{jk}(\tfrac{n+1}{2},m,l)   \,\,   \right].  \\   \nonumber
\end{eqnarray}
Notice that terms like ${\rm e}^{i\pi(j+m)}$ appearing in (\ref{sum4.1}) 
gave rise to $-{\rm e}^{i\pi j}$ in (\ref{term3}), since the contribuiting $m$'s are 
necessarilly odd according to (\ref{term2}).  
Also, according to (\ref{term2}), one can see that for $j+k$ odd (even), 
the above sum is purely imaginary (real). 
The conclusions from (\ref{sum3}) are also applied to ${{\bf I}_\star}_{jk}'$ in 
(\ref{sum4}). This completes the proof for $n$ {\it odd}.



\begin{thebibliography}{99}
\bibitem{endnote2}
To avoid misunderstandings we stress that 
$\lambda_m$ and $\mu_m$ are, respectively, 
a $m^{\text{th}}$ vector and a $m^{\text{th}}$ scalar of a set 
indexed by the labels of the $m^{\text{th}}$ Lindblad operator.
\bibitem{endnote3}
Note that the sum on Eq.(\ref{lindblad}) now encompasses 
the primed and the unprimed Lindblad operators.
\end{thebibliography}
\end{document}